\documentclass[aps,prd,showpacs,preprintnumbers]{revtex4}
\usepackage{graphicx}                                 
\usepackage{amsmath,amsfonts}

\usepackage{epsfig}
\usepackage{amssymb}
\usepackage{amsfonts}
\usepackage{float}
\newcommand{\beq}{\begin{equation}}
\newcommand{\eeq}{\end{equation}}
\newcommand{\bea}{\begin{eqnarray}}
\newcommand{\eea}{\end{eqnarray}}
\newcommand{\beas}{\begin{eqnarray*}}
\newcommand{\eeas}{\end{eqnarray*}}
\newcommand{\eq}{\begin{equation}}
\newcommand{\en}{\end{equation}}
\newcommand{\eqa}{\begin{eqnarray}}
\newcommand{\ena}{\end{eqnarray}}
\newcommand{\Fig}[1]{Fig.~\ref{#1}}
\newcommand{\Tab}[1]{Table~\ref{#1}}

\newcommand{\Eq}[1]{Eq.~(\ref{#1})}

\begin{document}

\preprint{ADP-07-20/T660, HU-EP-07/62, LMU-ASC 73/07, DESY 07-219}

\title{Vacuum structure revealed by over-improved stout-link smearing \\
  compared with the overlap analysis for quenched QCD}
\author{E.-M. Ilgenfritz\footnote{leave of absence from
    Humboldt-Universit\"at zu Berlin, Institut f\"ur Physik, 12489
    Berlin, Germany}}
\affiliation{Special Research Centre for the Subatomic Structure of
  Matter (CSSM), Department of Physics, University of Adelaide, Adelaide SA 5005,
  Australia}
\author{D.~Leinweber} 
\affiliation{Special Research Centre for the Subatomic Structure of
  Matter (CSSM), Department of Physics, University of Adelaide, Adelaide SA 5005,
  Australia}
\author{P.~Moran} 
\affiliation{Special Research Centre for the Subatomic Structure of
  Matter (CSSM), Department of Physics, University of Adelaide, Adelaide SA 5005,
  Australia}
\author{K.~Koller} 
\affiliation{Sektion Physik, Universit\"at M\"unchen, 80333 M\"unchen, Germany}
\author{G.~Schierholz} 
\affiliation{John von Neumann-Institut f\"ur Computing NIC, 15738 Zeuthen, Germany} 
\affiliation{Deutsches Elektronen-Synchrotron DESY, 22603 Hamburg, Germany}
\author{V.~Weinberg} 
\affiliation{John von Neumann-Institut f\"ur Computing NIC, 15738 Zeuthen, Germany} 
\affiliation{Institut f\"ur theoretische Physik, Freie Universit\"at Berlin, 
14196 Berlin, Germany}

\date{January 11, 2008}

\begin{abstract}
A detailed comparison is made between the topological structure of quenched
QCD as revealed by the recently proposed over-improved
stout-link
smearing in conjunction with an improved gluonic definition
of the topological density on one hand and a similar analysis made
possible by the overlap-fermionic topological charge density
both with and without
variable ultraviolet cutoff $\lambda_{cut}$. The matching is
twofold, provided by fitting the density-density two-point functions
on one hand and by a point-by-point fitting of the topological
densities according to the two methods.  We point out the similar
cluster structure of the topological density for moderate smearing and
$200 \mathrm{~MeV} < \lambda_{cut} < 600 \mathrm{~MeV}$, respectively.
We demonstrate the relation of the gluonic topological density for
extensive smearing to the location of the overlap zero modes and the
lowest overlap non-zero mode as found for the unsmeared
configurations.
\end{abstract}


\maketitle

\section{Introduction}
\label{sec:introduction}

Massless overlap fermions~\cite{Neuberger:1997fp,Neuberger:1998wv}
have provided us with a fermionic definition of topological
charge~\cite{Hasenfratz:1998ri,Niedermayer:1998bi}.  This offers the
advantage that, by truncating to the overlap modes with $|\lambda| <
\lambda_{cut}$, the effect of ultraviolet filtering can be
studied~\cite{Horvath:2002yn} without changing the gauge field itself.
A broad study of different aspects of vacuum structure, so far for
quenched QCD, has been published recently~\cite{Ilgenfritz:2007xu}.
On the other hand, during the 80's procedures of cooling or smearing of
gauge fields were proposed to exhibit the infrared structure of
gauge fields~\cite{Teper:1985ek,Ilgenfritz:1985dz}.  This has been
considered with reservations because it is difficult to assess in
which aspect the gauge field configuration could have changed under
this procedure. The practitioners of cooling/smearing, in particular
those who were focussing on vacuum structure in the form of extended,
smooth (semiclassical) structures, have continued to improve their
techniques: cooling with improved actions~\cite{Forcrand:1997sq},
restricted improved cooling~\cite{GarciaPerez:1998ru} etc).
Over-improved cooling~\cite{Perez:1993ki,Bruckmann:2004ib} has been
applied in order to prevent instantons or other topological
excitations from collapsing.

More recently, the concept of cooling/smearing has been modified in
another direction, to make it differentiable with respect to the
unsmeared field.  The so-called stout-link
smearing~\cite{Morningstar:2003gk} makes it possible to evaluate the
``force'' for the molecular dynamics in non-quenched update algorithms
with a fermionic action involving some form of ``fat'' (smeared) links
substituting the original gauge links. The concept of over-improvement
has also recently been applied to stout-link
smearing~\cite{Moran:2008ra}.

In this paper, following Ref.~\cite{Moran:2007nc} we want to
characterize the subsequent mapping ``link $\rightarrow$ stout link''
as a particular relaxation scheme, eventually leading to a finally
classical configuration.  One obvious way to discuss this process is
to record the local distribution of topological charge density
expressed by an improved~\cite{BilsonThompson:2002jk} gluonic
definition of field strength and topological density. In this paper we
shall compare the emergent structures with the topological density
provided by the overlap definition~\cite{Ilgenfritz:2007xu} with
different levels of ultraviolet filtering. Surprisingly, the overlap
topological charge density without filtering, that recently has been
found to form lower-dimensional
structures~\cite{Horvath:2003yj,Ilgenfritz:2007xu,Ilgenfritz:2007ua},
corresponds to few steps of smearing. This comparison will lead us to
a one-to-one mapping of the ultraviolet cutoff $\lambda_{cut}$ (mode
truncation in the overlap picture) to the number of stout-link
smearing iterations over a wide range of smearing iterations. In a
similar spirit, the correspondence of APE
smearing~\cite{DeGrand:2000gq}, Laplacian
filtering~\cite{Bruckmann:2005hy} and the topological density filtered
according to another Ginsparg-Wilson Dirac
operator~\cite{Gattringer:2000qu,Gattringer:2000js} has been studied
recently~\cite{Bruckmann:2006wf,Solbrig:2007nr}.

Only a few iterations of stout-link smearing are necessary before
structures become recognizable with the gluonic definition of the
topological density, and these structures are surprisingly far 
from 4D extended, sign-coherent lumps, such that the topological
density compares well with the unfiltered overlap definition. 
We have two criteria to establish this relation
between smearing and filtering.  First, it is the behavior of the
two-point correlation function of the topological density that emerges
from the respective definition. The second is the actual site-by-site
difference of the topological density profile over a set of lattice
configurations. The quality of the latter coincidence is surprisingly
good, which supports the reliability of both methods to explore the
vacuum structure.

Finally, however, this relation becomes loose because stout-link
smearing turns the configurations into piecewise classical fields
which apparently resemble instantons and anti-instantons. Although the
coherence among the lowest overlap modes guarantees a relatively
simple picture of the fermionic topological density and the
ultraviolet filtered gluonic field strength with low cutoff
$\lambda_{cut}$~\cite{Ilgenfritz:2007xu}, there is
no argument as to why the
overlap picture should be in correspondence to a (linkwise) classical
lattice configuration.

It is intriguing to see that the instanton-like structure that is
revealed in this late stage of smearing corresponds to the overlap
zero mode(s) and the lowest pair(s) of overlap non-zeromodes obtained
for the respective unsmeared (equilibrium) configuration. A similar
observation has already been made by 
Negele {\it et al.}~\cite{Ivanenko:1997nb}. In Ref.~\cite{Ilgenfritz:2007ua} some
of us have presented a cluster analysis of individual eigenmodes of
the overlap Dirac operator. It turns out that the moderate number of
clusters that the zero and first non-zero modes consist of (at a level
of scalar density below the peak values) are pointing towards the
positions where instantons and anti-instantons appear later.

Our paper is organized as follows: in Section \ref{sec:technicalities}
we explain briefly the overlap operator and the stout-link smearing
procedure, in Section \ref{sec:matching} we describe the matching
between smearing and ultraviolet filtering, according to the two-point
correlator and according to a global fitting of the profile of charge.
In Section \ref{sec:cluster} we try to relate the clusters that both
definitions exhibit to each other.
In the final stadium of smearing
we shall confront the emergent semiclassical lumps with the lowest
eigenmodes (zero mode and lowest non-zero mode) of the original
configurations. In Section \ref{sec:conclusions} we draw conclusions
and give an outlook.

\section{Technical Details}
\label{sec:technicalities}

\subsection{Configurations and the overlap definition of topological
  density}
\label{subsec:overlapdefinition}

The configurations underlying this comparison stem from an extended
investigation published in Ref.~\cite{Ilgenfritz:2007xu}. The
configurations are taken from an ensemble of $16^3\times32$ quenched
lattices generated with the 
tadpole improved L\"uscher-Weisz
action at $\beta=8.45$.

The (massless) overlap Dirac operator is constructed for the
Wilson-Dirac input kernel $D_W = M - \rho/a$, 
$M$ being the massless Wilson-Dirac operator with $r=1$, and $\rho=1.4$.
The corresponding
solution of the Ginsparg-Wilson relation reads as follows
\begin{equation}
  D(0) = \frac{\rho}{a}\left( 1 + D_W/\sqrt{D_W^{\dagger} D_W} \right) =  \frac{\rho}{a}\left( 1 + \gamma_5 \mathrm{sgn} (H_W) \right) \; ,
  \label{eq:overlap}
\end{equation}
with $H_W = \gamma_5 D_W$. Circa 150 overlap eigenmodes have been
obtained per configuration. They have been used to construct
ultraviolet smeared topological densities according to a cut-off
$\lambda_{cut}=200 \mbox{~MeV}$, $400 \mbox{~MeV}$ and $635
\mbox{~MeV}$.  For half of the subset of 10 configurations
particularly considered in the present study we have also calculated
for Refs.~\cite{Ilgenfritz:2007xu,Ilgenfritz:2007ua} the overlap 
topological density without mode truncation, the ``all-scale'' 
topological density.

The spectrum eventually consists of some zero modes, 
in addition to pairs of non-zero modes of globally vanishing chirality.  
The topological density can be formally obtained from the trace of the
overlap Dirac operator~\cite{Niedermayer:1998bi}
\begin{equation}
  q(x) = - \mathrm{tr} \left[ \gamma_5 \left( 1 - \frac{a}{2} D(0;x,x) \right) \right] \; .
  \label{eq:full_density}
\end{equation}
Using the spectral representation of the overlap Dirac operator, a
family of ultraviolet filtered topological charge densities labelled
by $q_{\lambda_{cut}}(x)$ can be
obtained~\cite{Horvath:2002yn,Koma:2005sw},
\begin{equation}
  q_{\lambda_{cut}}(x) = - \sum_{|\lambda|<\lambda_{cut}} 
  \left( 1 - \frac{\lambda}{2} \right) \psi_{\lambda}^{\dagger}(x) \gamma_5 \psi_{\lambda}(x) \; .
  \label{eq:truncated_density}
\end{equation}
The topological charge of each configuration
fulfills the Atiyah-Singer index theorem~\cite{Hasenfratz:1998ri}
\begin{equation}
  Q = n_{-} - n_{+} \; .
  \label{eq:Atiyah-Singer}
\end{equation} 
Note that the zero modes of any given configuration carry the same
chirality.

In this paper we shall point out that 
the above-mentioned  family of densities is well
represented by the gluonic topological density after an appropriate
number of iterations 
of stout-link smearing.

\subsection{Over-improved stout-link smearing}
\label{subsec:smearing}

In Refs.~\cite{Moran:2008ra,Moran:2007nc} two of us have motivated the
stout-link smearing with respect to a specific over-improved
action. The concept of over-improved action arose from the observation
that the Wilson one-plaquette action does not guarantee the stability
of instantons~\cite{Gockeler:1989qg}.  Even with the Symanzik improved
action including $1\times1$ plaquettes and $1 \times 2$ rectangular
loops, for a classical instanton solution of size $\rho_{inst}$ placed
on the lattice, the action is obtained as
\begin{equation}
  S^{inst} = \frac{8\pi^2}{g^2} 
  \left[ 1 - \frac{17}{210}\left(\frac{a}{\rho_{inst}}\right)^4 \right] \; .
  \label{eq:instanton_without_epsilon}
\end{equation}
Following Garcia Perez {\it et al.}~\cite{Perez:1993ki},
Ref.~\cite{Moran:2008ra} proposes to modify the relative
contribution of plaquette and rectangular terms, parametrized by a
parameter $\epsilon$. For the Symanzik action this reads as follows
\begin{equation}
  S(\epsilon) = \beta \sum_x \sum_{\mu < \nu} \left[\frac{5-2\epsilon}{3}~(1 - P_{\mu\nu}(x))
    - \frac{1-\epsilon}{12}\left(1 - R_{\mu\nu}(x) + 1 - R_{\nu\mu}(x) \right) \right] \; ,
  \label{eq:overimproved_Symanzik}
\end{equation}
interpolating between Wilson ($\epsilon=1$) and Symanzik action
($\epsilon=0$).  As usual, $\beta=6/g^2$. 
The plaquettes are denoted as 
$$P_{\mu\nu} = \frac{1}{3}
\mathrm{Re~Tr}~\left(U_{x,\mu}
                     U_{x+\hat{\mu},\nu}
	   	     U^{\dagger}_{x+\hat{\nu},\mu}
		     U^{\dagger}_{x,\nu}\right)$$ 
and the traces of the rectangular loops as 
$$R_{\mu\nu} = \frac{1}{3}
\mathrm{Re~Tr}~\left(U_{x,\mu}
                     U_{x+\hat{\mu},\mu}
		     U_{x+2\hat{\mu},\nu}
		     U^{\dagger}_{x+\hat{\nu}+\hat{mu},\mu} 
		     U^{\dagger}_{x+\hat{\nu},\mu} 
		     U^{\dagger}_{x,\nu}\right) \; . $$ 
With this action the instanton action behaves as
\begin{equation}
  S^{inst} = \frac{8\pi^2}{g^2} \left[ 1 - \frac{\epsilon}{5} 
  \left(\frac{a}{\rho_{inst}}\right)^2
    +\frac{14\epsilon-17}{210}\left(\frac{a}{\rho_{inst}}\right)^4 \right] \; .
  \label{eq:instanton_with_epsilon} 
\end{equation}
In Ref.~\cite{Moran:2007nc} a value $\epsilon=-0.25$ was found to provide
an instanton action almost independent of the radius for
$\rho_{inst} > 1.5 a$.

The stout-link iteration solves the equation of motion (without momentum)
\begin{equation}
U^{old}_{x\mu} \to U^{new}_{x\mu} =
\exp\left( - \rho_{sm}~\frac{\delta S(\epsilon)} {\delta U_{x\mu}} \right)~U^{old}_{x\mu}
\label{eq:stout_link_iteration}
\end{equation}
in parallel, with a ``time step'' $\rho_{sm}$.
No projection onto the group $SU(3)$ is required in distinction to APE
smearing.
The derivative is the traceless antihermitean part of a sum of untraced loops
$$\frac{\delta S(\epsilon)}{\delta U_{x\mu}}=
\frac{5 - 2\epsilon}{3} \sum \left(\mathrm{1x1~loops~touching~} U \right)
+ \frac{1 - \epsilon}{12} \sum \left(\mathrm{1x2+2x1~loops~touching~} U \right)_{|\mathrm{antiherm~traceless}}$$
with $U_{x\mu}$ in the leftmost position.
This method has been discussed in Ref.~\cite{Moran:2007nc} 
with respect to the two-point
correlator of the topological density. The differences between
quenched and dynamical ensembles of configurations with respect to the
distribution of topological density have been demonstrated. The
instanton-like features are seen to become stronger with more and more
iterations of (\ref{eq:stout_link_iteration}).
The resemblance to instantons was stated~\cite{Moran:2007nc} by a close 
correlation --
maximum by maximum of the modulus of the topological density -- 
between the ``instanton size'', determined by the curvature of $|q(x)|$
in the maximum position $x_0$ of the cluster, and the value of the 
density $|q(x_0)|$.

The local lattice operator~\cite{BilsonThompson:2002jk} to represent
the topological density for the stout-link smeared configurations is
based on a highly improved field-strength tensor,
\begin{equation}
  F^{imp}_{\mu\nu}(x) 
  = k_1 C^{(1,1)}_{\mu\nu}(x) + k_2 C^{(2,2)}_{\mu\nu}(x) 
  + k_3 C^{(1,2)}_{\mu\nu}(x) + k_4 C^{(1,3)}_{\mu\nu}(x) 
  + k_5 C^{(3,3)}_{\mu\nu}(x)
\end{equation}
with
\begin{eqnarray}
  k_1 & = & \frac{19}{9} - 55 k_5 \; , \nonumber \\
  k_2 & = & \frac{1}{36} - 16 k_5 \; , \nonumber \\
  k_3 & = & -\frac{32}{45} + 64 k_5 \; , \nonumber \\
  k_4 & = & \frac{1}{15} - 6 k_5 \; , \nonumber \\
\end{eqnarray}
and with
\begin{equation}
  C^{(nm)}_{\mu\nu}(x) = \frac{1}{8} \left( W^{(n,m)}_{\mu\nu}(x) 
    + W^{(m,n)}_{\mu\nu}(x) \right)
\end{equation}
being a symmetrized ``clover sum'' of
$(n\times\hat{\mu},m\times\hat{\nu})$ Wilson loops around the site
$x$. A 3-loop improved field strength tensor can be achieved chosing
$k_5=1/90$, such that $k_3 = k_4 = 0$.  The topological charge density
is then represented in the form
\begin{equation}
  q_{sm}(x) =  \frac{g^2}{16\pi^2} \mathrm{Tr}~\left( F_{\mu\nu}\tilde{F}_{\mu\nu}\right) \; .
\label{eq:gluonic_density}
\end{equation}

\section{Matching stout-link smearing to overlap filtering}
\label{sec:matching}
 
\subsection{Matching the two-point correlator}
\label{subsec:correlator}

The two-point correlators for stout-link smearing and overlap
filtering are matched using a minimization of the sum of the absolute
difference between the correlators. That is, we compute
\begin{equation}
  \mathrm{min}\,\left( \sum_x |\langle q(x)q(0) \rangle_{sm} - \langle
    q(x)q(0) \rangle_{\lambda_{cut}}| \right) 
  \label{eq:matchqxqy}
\end{equation}
as a function of the number of smearing sweeps, $n_{sw}$,
for fixed $\lambda_{cut}$.

The topological charge of each overlap filtered configuration is
integer valued because they satisfy the Atiyah-Singer index
theorem~(\ref{eq:Atiyah-Singer}). For stout-link smearing it can take
up to $10$ sweeps of smearing to achieve an integer charge on these
$16^3\times32$ lattices. In order to 
compare overlap filtering and stout-link smearing for $n_{sw} < 10$ 
in a fair way a non-perturbative
normalization is applied. Given that we know the topological charge $Q$
from the overlap configurations, we calculate, for each 
number of smearing sweeps $n_{sw}$, a normalization factor $Z_{sm}$ via
\begin{equation}
  Q = Z_{sm} \sum_x q_{sm}(x) \,,
  \label{eq:zswnormalization}
\end{equation}
and $q_{sm}(x)$ is then normalized through $q_{sm}(x) \rightarrow
Z_{sm}\, q_{sm}(x)$. This ensures that $Q_{sm} = Q$. We also experimented
with an alternate normalization where we matched the absolute values of 
the topological charge density, however this 
proved less fruitful.

Typical values of $Z_{sm}$ are provided in
\Tab{table:Zfactors}.
\begin{table}
  \begin{center}
    \begin{tabular}{c@{\hspace{10pt}}c@{\hspace{10pt}}c@{\hspace{10pt}}c@{\hspace{10pt}}c@{\hspace{10pt}}c}
      \hline
      \hline
      $n_{sw}$ & $Q = -8$ & $Q = 0$ & $Q = -1$ & $Q = -7$ & $Q = 0$ \\
      \hline
      \hline
      1 &   0.9954 &  -0.0001 &   1.0054 &   0.6442 &   0.0000 \\
      2 &   0.9634 &  -0.0001 & -21.5369 &   0.7987 &  -0.0001 \\
      3 &   0.9494 &  -0.0001 &   3.6497 &   0.8990 &  -0.0001 \\
      4 &   0.9585 &  -0.0002 &   1.8690 &   0.9485 &  -0.0001 \\
      5 &   0.9760 &  -0.0004 &   1.4537 &   0.9728 &  -0.0001 \\
      6 &   0.9903 &  -0.0008 &   1.2736 &   0.9843 &  -0.0002 \\
      7 &   0.9982 &  -0.0015 &   1.1725 &   0.9894 &  -0.0004 \\
      8 &   1.0010 &  -0.0030 &   1.1105 &   0.9915 &  -0.0006 \\
      \hline
    \end{tabular}
  \end{center}
  \caption{Table of $Z_{sm}$ values for the initial five
    configurations. The normalization procedure is only valid for $Q
    \ne 0$ and works best for large $|Q|$. The fluctuating
    values for the configuration with $Q=-1$ occuring at
    $n_{sw} = 2$ are due to $Q_{sm}$ being approximately $0$ at this point.}
  \label{table:Zfactors}
\end{table}
For obvious reasons, it is only possible to extract a $Z_{sm}$ factor
for $Q \ne 0$, and the procedure works best for $Q$ far from
zero. Consequently, the best results are found for the configurations
where $Q = -7$ and $-8$, where the $Z_{sm}$ values rapidly approach
1. This is also the case for the configuration with $Q=-1$, however
$Z_{sm}$ fluctuates at $n_{sw} = 2$, which is due to $Q_{sm}$ being
approximately $0$ at this point.

\begin{table}
  \begin{center}
    \begin{tabular}{c@{\hspace{20pt}}cc@{\hspace{20pt}}cc@{\hspace{20pt}}cc@{\hspace{20pt}}cc@{\hspace{20pt}}cc}
      \hline
      \hline
      $\lambda_{cut}$ & \multicolumn{2}{l}{$Q = -8$}&\multicolumn{2}{l}{$Q = 0$}&\multicolumn{2}{l}{$Q = -1$}&\multicolumn{2}{l}{$Q = -7$}&\multicolumn{2}{l}{$Q = 0$}\\
      \hline
      \hline
      full density & 005 & 005 & 005 & --- & 005 & 005 & 005 & 004 & 005 & ---\\
      all known modes & 039 & 039 & 043 & --- & 036 & 036 & 038 & 037 & 034 & ---\\
      634 MeV      & 052 & 052 & 064 & --- & 048 & 048 & 055 & 055 & 048 & ---\\
      400 MeV      & 127 & 127 & 183 & --- & 109 & 108 & 139 & 138 & 115 & ---\\
      200 MeV      & 248 & 247 & 300 & --- & 232 & 232 & 300 & 300 & 271 & ---\\
      \hline
    \end{tabular}
  \end{center}
  \caption{The best matches for the two-point correlators of five
    different configurations as determined by \Eq{eq:matchqxqy}. For
    each configuration, the two columns give the number of smearing
    sweeps that correspond to the best match for a given level of UV
    filtering. In each case the left column gives the best match for
    the unnormalised gluonic density and the right column gives the
    best match for the gluonic density normalised with $Z_{sm}$ (see
    text). The right columns for the $Q=0$ configurations are absent
    because the normalization procedure fails in these cases.}
  \label{table:match-correl}
\end{table}
In Table~\ref{table:match-correl} we present the best matches for the
two-point correlators between the filtered overlap densities and the
stout-link smeared gluonic densities. Configurations have been used
for which the unfiltered overlap topological density (full density)
has been measured. The correlators of the fermionic topological density 
including
all known modes, with $\lambda_{cut}=634 \mathrm{~MeV}$, with
$\lambda_{cut}=400 \mathrm{~MeV}$ and with $\lambda_{cut}=200
\mathrm{~MeV}$ are matched against the smeared gluonic correlators.

\begin{figure*}
  \vspace*{0.8cm}
  \includegraphics[width=0.44\textwidth,angle=90]{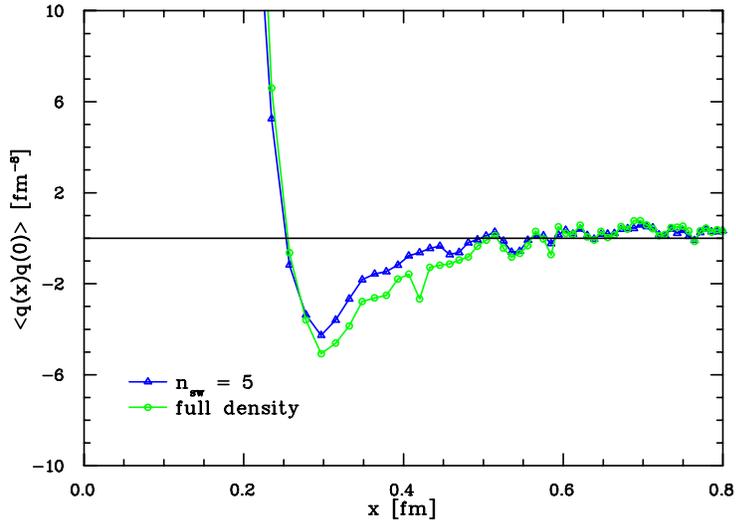}
  \caption{The two-point function of the fermionic topological density
    for one configuration shown without mode truncation compared with
    the bosonic definition after 5 steps of smearing.}
  \label{fig:Fig1}
\end{figure*}
\begin{figure*}[!h]
  \vspace*{0.8cm}
  \includegraphics[width=0.44\textwidth,angle=90]{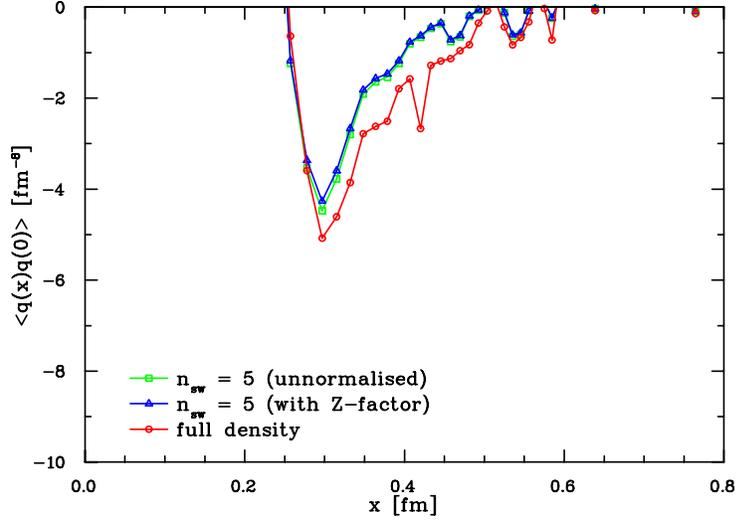}
  \caption{The same as in Fig. 1 with the effect of the
    renormalization factor $Z_{sm}=Q_{ferm}/Q_{sm}$ also shown.
    ($Q_{ferm}=8$, $Q_{sm}=8.19$)}
  \label{fig:Fig2}
\end{figure*}
\begin{figure*}[!h]
  \vspace*{0.8cm}
  \includegraphics[width=0.44\textwidth,angle=90]{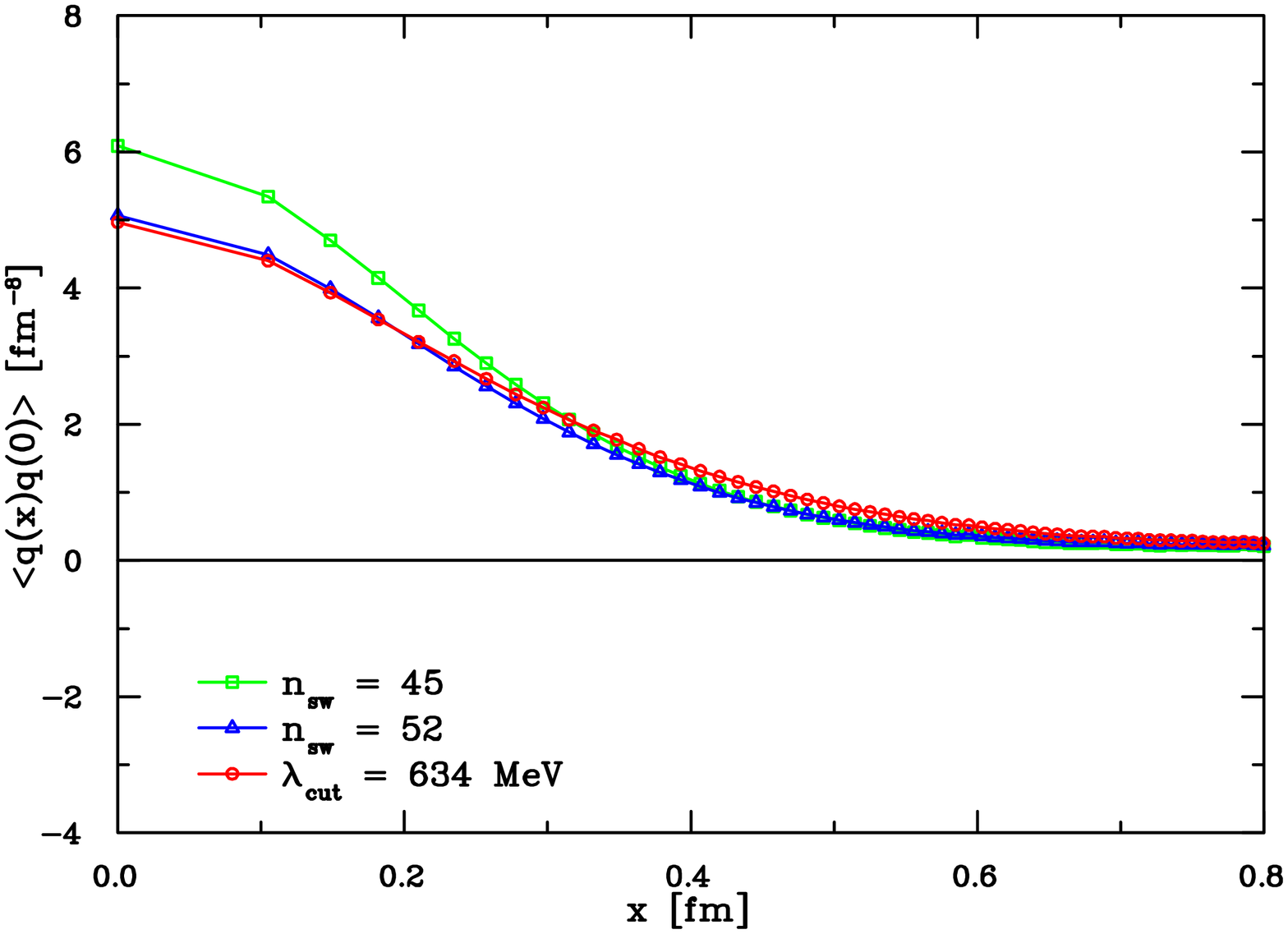}
  \caption{The two-point function of the fermionic topological density
    with an UV cutoff at $\lambda_{cut}=634 \mathrm{~MeV}$ compared
    with the bosonic definition after 52 steps of smearing when the
    correlator is fitted best. Smearing after 45 steps leads to
    the best global matching of the density $q_{\lambda_{cut}}$ with
    $\lambda_{cut}=634 \mathrm{~MeV}$, but the correlator is steeper.}
  \label{fig:Fig3}
\end{figure*}
We show in Fig. 1 the two-point correlator of the topological density
for a single configuration represented by the unfiltered fermionic
topological density (full density) of Eq.~(\ref{eq:full_density}) of the
equilibrium configuration on one hand and for the gluonic definition
of Eq.~(\ref{eq:gluonic_density}) after 5 smearing steps. It is
remarkable that the two correlators follow each other's fluctuations
at larger distance. There is some difference in normalization of the
negative peak. This configuration has $Q_{ferm}=-8$, and the gluonic
definition gives $Q_{sm}=-8.19$.  In this case the
``non-perturbative'' renormalization would even slightly increase the
difference between the curves 
as the three curves in Fig. 2 show.

In Fig.~3 the density-density two-point correlator for the same
configuration is compared for an ultraviolet cutoff $\lambda_{cut}=634
\mathrm{~MeV}$ for the fermionic, overlap definition and the gluonic
definition after 52 smearing steps. The filtered correlators match
perfectly.  The third curve shows the correlator with the gluonic
definition after 45 smearing steps.  This refers to the case of an
optimal point-by-point matching of the densities, as discussed in the
next section.  The corresponding correlator is somewhat higher and
steeper because it uses slightly less smearing iterations.

\begin{figure*}
  \vspace*{0.8cm}
  \includegraphics[width=0.44\textwidth,angle=90]{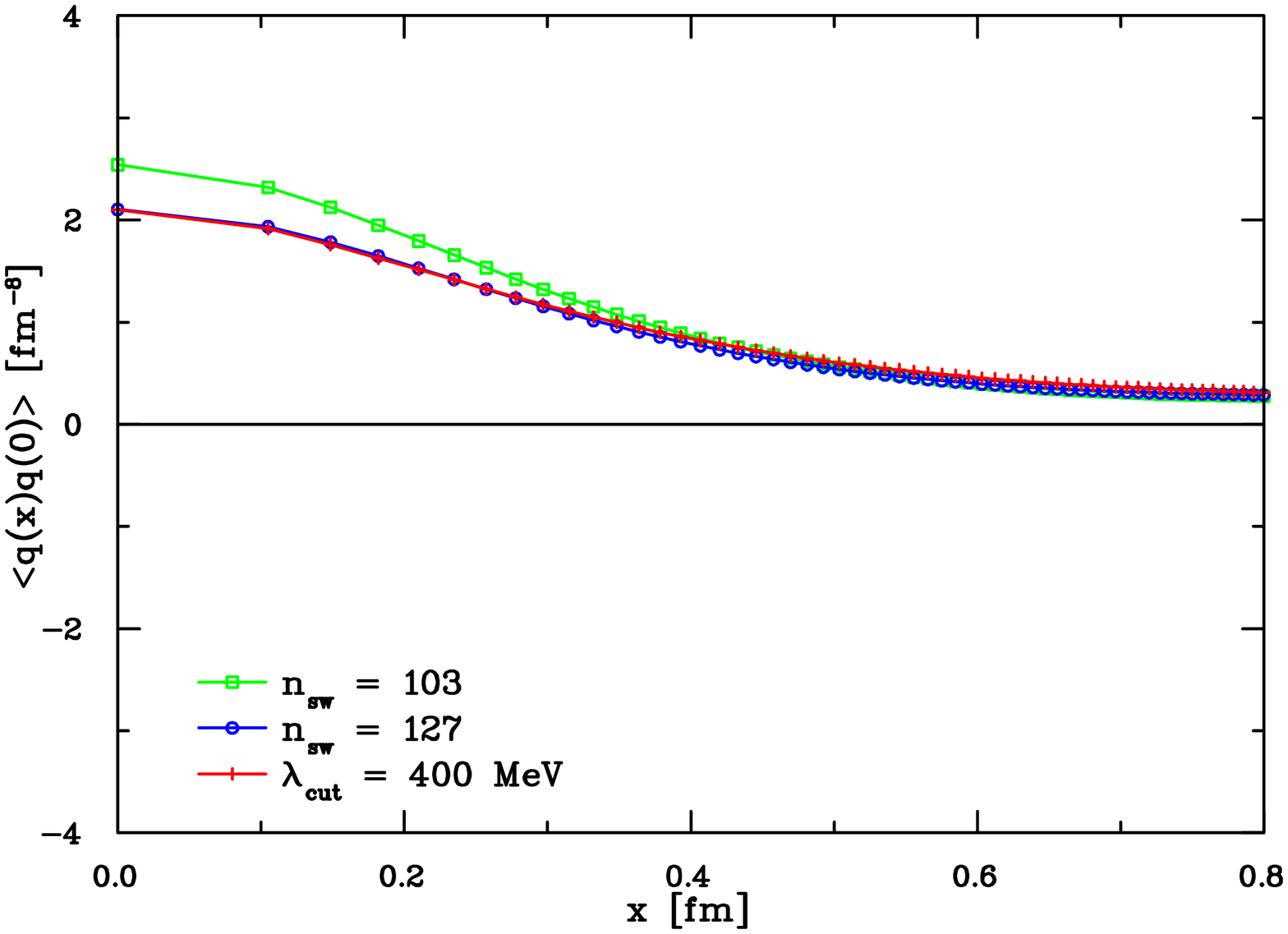}
  \caption{The same as Fig. 3 with an UV cutoff at $\lambda_{cut}=400
    \mathrm{~MeV}$ fitted best by the bosonic definition after 127
    steps of smearing. Smearing after 103 steps leads to the best
    global matching of the densities, but the correlator is steeper.}
  \label{fig:Fig4}
\end{figure*}
Fig.~4 shows the same configuration for a lower ultraviolet cutoff
$\lambda_{cut}=400 \mathrm{~MeV}$.  After 127 smearing steps the
correlator for the gluonic definition fits the correlation function
perfectly. Somewhat less smearing steps (103), again optimally fitting the
fermionic topological density point by point, tend to overestimate the correlator.

\begin{figure*}
  \vspace*{0.8cm}
  \includegraphics[width=0.44\textwidth,angle=90]{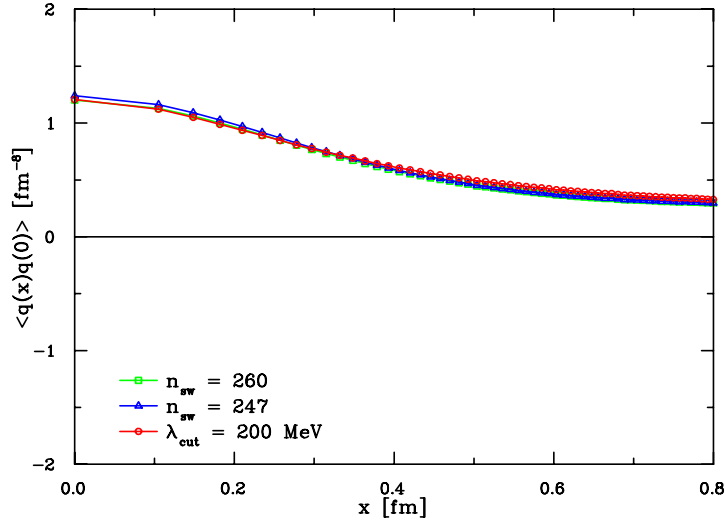}
  \caption{The same as Figs. 3 and 4, now with an UV cutoff at
    $\lambda_{cut}=200 \mathrm{~MeV}$ fitted best by 248 sweeps of
    smearing. 260 sweeps of smearing gives the best match for the
    topological charge densities.}
  \label{fig:Fig5}
\end{figure*}

All smeared correlators discussed above were generated using a
smearing parameter of $\rho_{sm} = 0.06$. By using a different value for
$\rho_{sm}$ it is possible that a different number of smearing sweeps will
provide the best fit. This is because the amount of smearing applied
to a gauge field is proportional to $\rho_{sm}\,n_{sw}$. 
By varying $\rho_{sm}$
as well as the number of sweeps one has greater fine-grained control
over the matching. Using a variable $\rho_{sm}$, but holding $n_{sw} = 5$
fixed, and applying this to the same configuration considered previously 
we find that
$\rho_{sm} = 0.055$ provides the best match for the unfiltered topological
density.

\begin{figure*}
  \vspace*{0.8cm}
  \includegraphics[width=0.44\textwidth,angle=90]{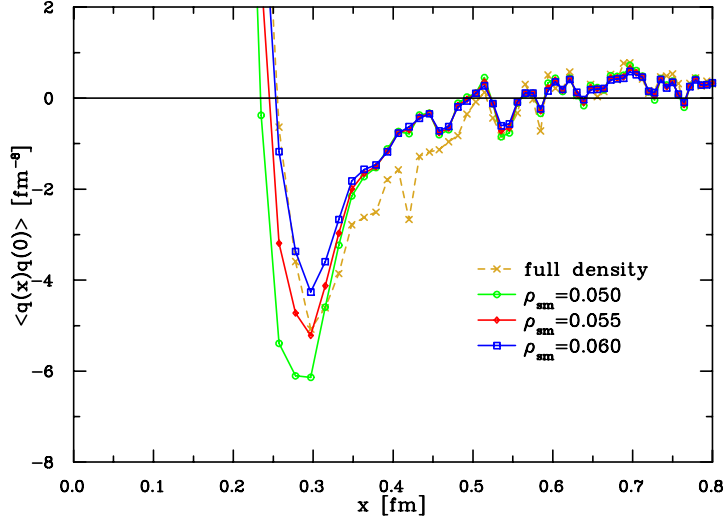}
  \caption{A comparison of the smeared two-point function for three
    different values of the smearing parameter, $\rho_{sm}$. The
    configuration shown is the same as that used in the previous
    figures. A best match as dictated by \Eq{eq:matchqxqy} is given by
    $\rho_{sm} = 0.055$, however $\rho_{sm} = 0.06$ gives the 
    best match for the $x$-intercept.}
  \label{fig:qxqy.r05.r06.r055}
\end{figure*}

A comparison of some different
values for for the smearing parameter is provided in
\Fig{fig:qxqy.r05.r06.r055}.
As expected, increasing $\rho_{sm}$ results in a suppression of the
negativity of the
two-point correlator. Although $\rho_{sm} = 0.055$ provides the best match
through a minimization of the sum of the differences in the two-point
functions, $\rho_{sm} = 0.06$ gives the best match for the $x$-intercept.
The two-point function for this larger $\rho_{sm}$ also appears to have a
similar shape to the two-point function of the full fermionic topological 
density.

\subsection{Matching the fermionic and gluonic topological density point by point}
\label{subsec:landscape}

We now aim to match the filtered fermionic densities with the smeared
gluonic densities through a point-by-point matching of the respective
topological charge densities $q_{\lambda_{cut}}(x)$ and $q_{sm}(x)$. 
Given some filtered
fermionic topological charge density we compare it against the gluonic density
$q_{sm}(x)$ after some number of sweeps by calculating the absolute
value of the difference between the respective $q$ at each $x$. To find the best
match we compute the minimum of the sum of the differences,
\begin{equation}
  \mathrm{min}\,\left( \sum_x | q_{sm}(x) - q_{\lambda_{cut}}(x) | \right) \; ,
  \label{eq:matchqx}
\end{equation}
as a function of the number of smearing sweeps, $n_{sw}$,
for fixed $\lambda_{cut}$.
The non-perturbative normalization of Eq.~(\ref{eq:zswnormalization})
will also be applied. The best matches are presented in
Table~\ref{table:match-x-by-x}.
\begin{table}
  \begin{center}

    \begin{tabular}{c@{\hspace{20pt}}cc@{\hspace{20pt}}cc@{\hspace{20pt}}cc@{\hspace{20pt}}cc@{\hspace{20pt}}cc}
      \hline
      \hline
      $\lambda_{cut}$ & \multicolumn{2}{l}{$Q = -8$}&\multicolumn{2}{l}{$Q = 0$}&\multicolumn{2}{l}{$Q = -1$}&\multicolumn{2}{l}{$Q = -7$}&\multicolumn{2}{l}{$Q = 0$}\\
      \hline
      \hline
      full density & 300 & 005 & 300 & --- & 300 & 006 & 300 & 005 & 300 & --- \\
      all known modes & 037 & 036 & 037 & --- & 035 & 035 & 035 & 035 & 033 & --- \\
      634 MeV      & 045 & 045 & 048 & --- & 044 & 044 & 048 & 048 & 044 & --- \\
      400 MeV      & 103 & 103 & 098 & --- & 103 & 103 & 087 & 087 & 093 & --- \\
      200 MeV      & 261 & 260 & 300 & --- & 236 & 236 & 192 & 191 & 187 & --- \\
      \hline
    \end{tabular}
  \end{center}
  \caption{Best matches for $q(x)$ between the
    filtered overlap densities and smeared gluonic densities. The left
    columns contain the best matches with the unnormalised gluonic
    densities, and the right columns contain the best matches when
    including the $Z_{sm}$ factor. We see the importance of the
    $Z_{sm}$ normalization factor when attempting the match the full
    fermionic density. The right columns are absent for the $Q=0$
    configurations because the normalization method does not work for
    this $Q$.}
  \label{table:match-x-by-x}
\end{table}
For a small number of sweeps the importance of the $Z_{sm}$
normalization is apparent. Visualizations of some configurations are
shown later in Section~\ref{sec:cluster}.

\subsection{Towards the no-smearing limit}
\label{subsec:nosmearing}

Increased smearing leads to flattening of the Euclidean two-point
function. It follows that less sweeps of smearing leads to 
an increasing
negative dip in the correlator, and we now study the behaviour of the
two-point function in the limit $n_{sw} \rightarrow 0$. After only one
or two sweeps of smearing there is a non-trivial renormalization that
must be applied to the topological charge density. From
\Tab{table:Zfactors} we see that after three smearing iterations the
$Z_{sm}$ factors for the $Q = -7$ and $Q = -8$ configurations are in
relatively good agreement. We can therefore study the correlator for
$n_{sw} \ge 3$, averaging over these two configurations.

\begin{figure*}[h]
  \begin{center}
    \includegraphics[width=0.44\textwidth,angle=90]{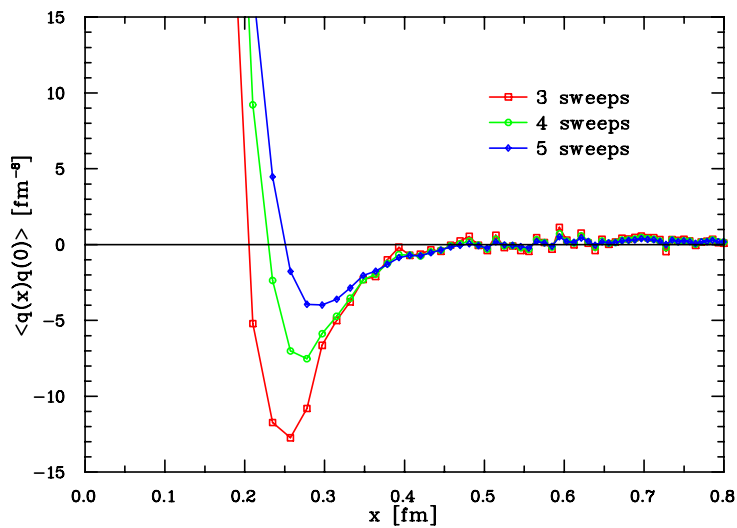}
    \caption{The two-point function as a function of smearing
      sweeps. The correlator has been averaged over the two
      configurations with $Q=-7$ and $Q=-8$. Both charge densities
      were normalised by an average of the $Z_{sm}$ values for the two
      configurations. As the number of sweeps is decreased the
      negative dip increases in magnitude and the $x$-intercept moves
      further toward zero.}
  \label{fig:qxqy.3.4.5.sweeps}

  \end{center}
\end{figure*}

Such a comparison leads to a series of correlators which
are displayed in Fig.~\ref{fig:qxqy.3.4.5.sweeps}.
The $x$-intercept also moves further towards zero.

\section{Comparison of topological clusters}
\label{sec:cluster}

\subsection{Clusters of both topological densities compared for weak
  stout-link smearing}
\label{subsec:cluster1}

\begin{figure*}[!h]
  \begin{center}
      \includegraphics[width=0.369\textwidth,angle=90]{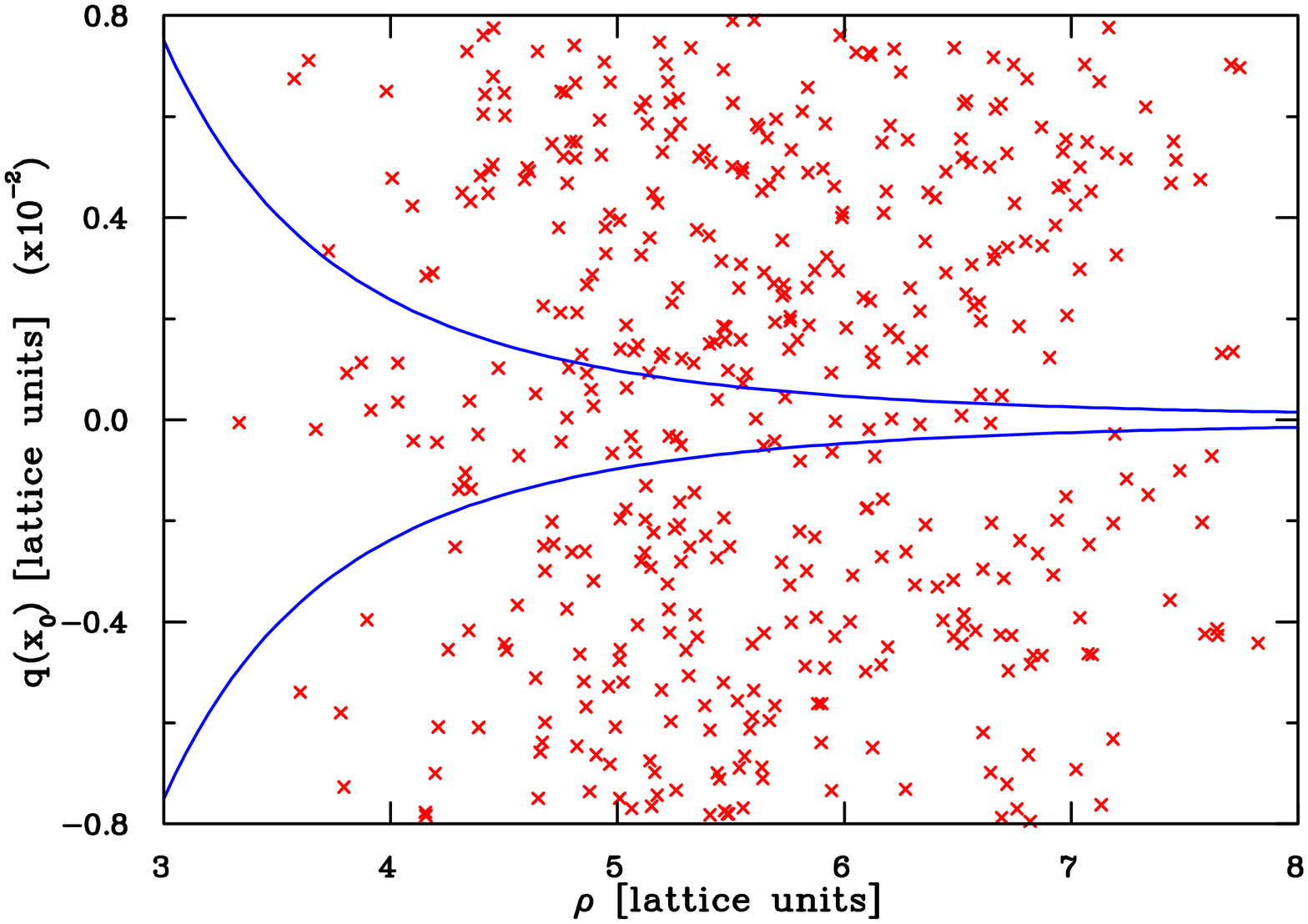} \\
      \vspace{0.5cm}
      \includegraphics[width=0.369\textwidth,angle=90]{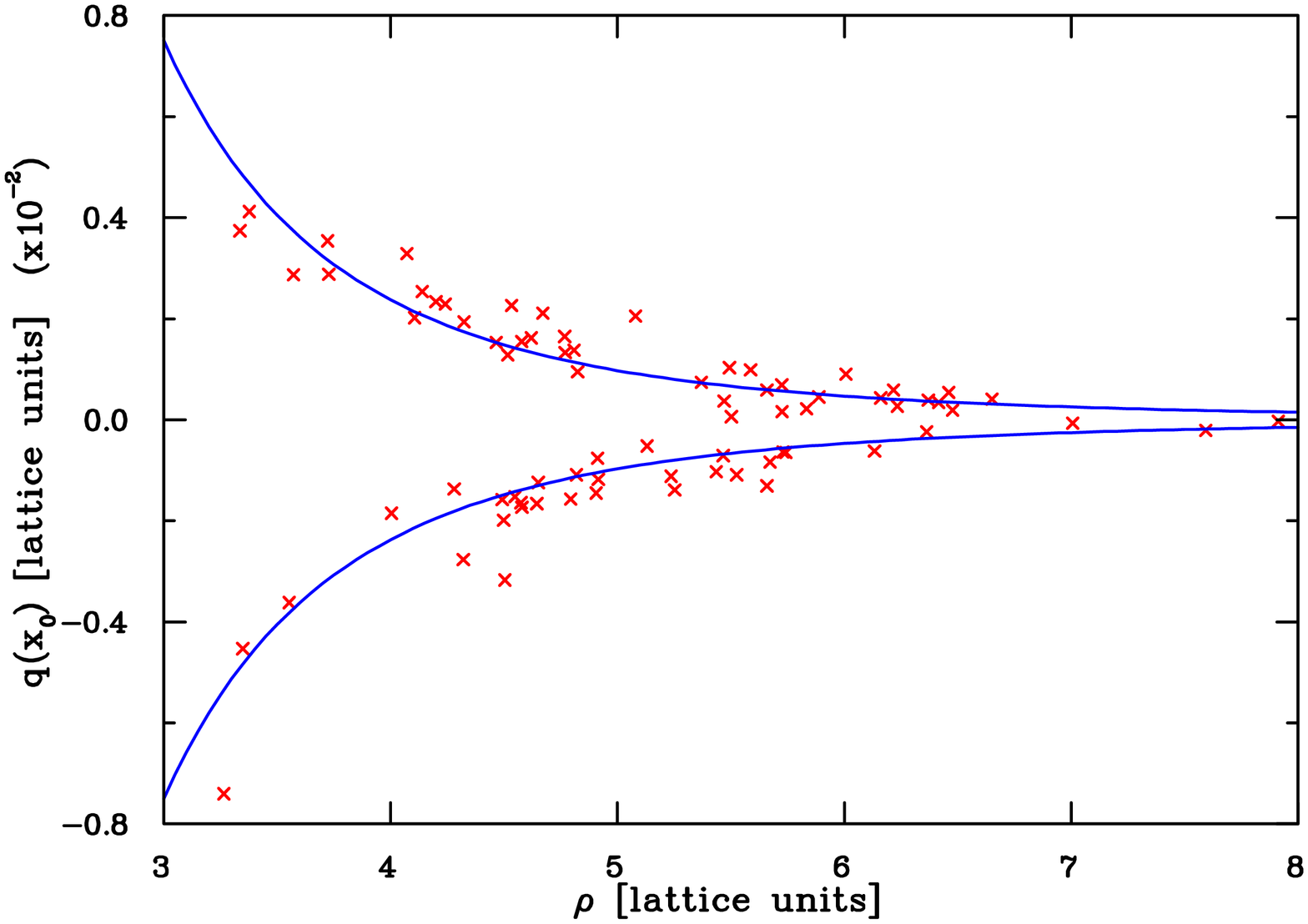} \\ 
      \vspace{0.5cm}
      \includegraphics[width=0.369\textwidth,angle=90]{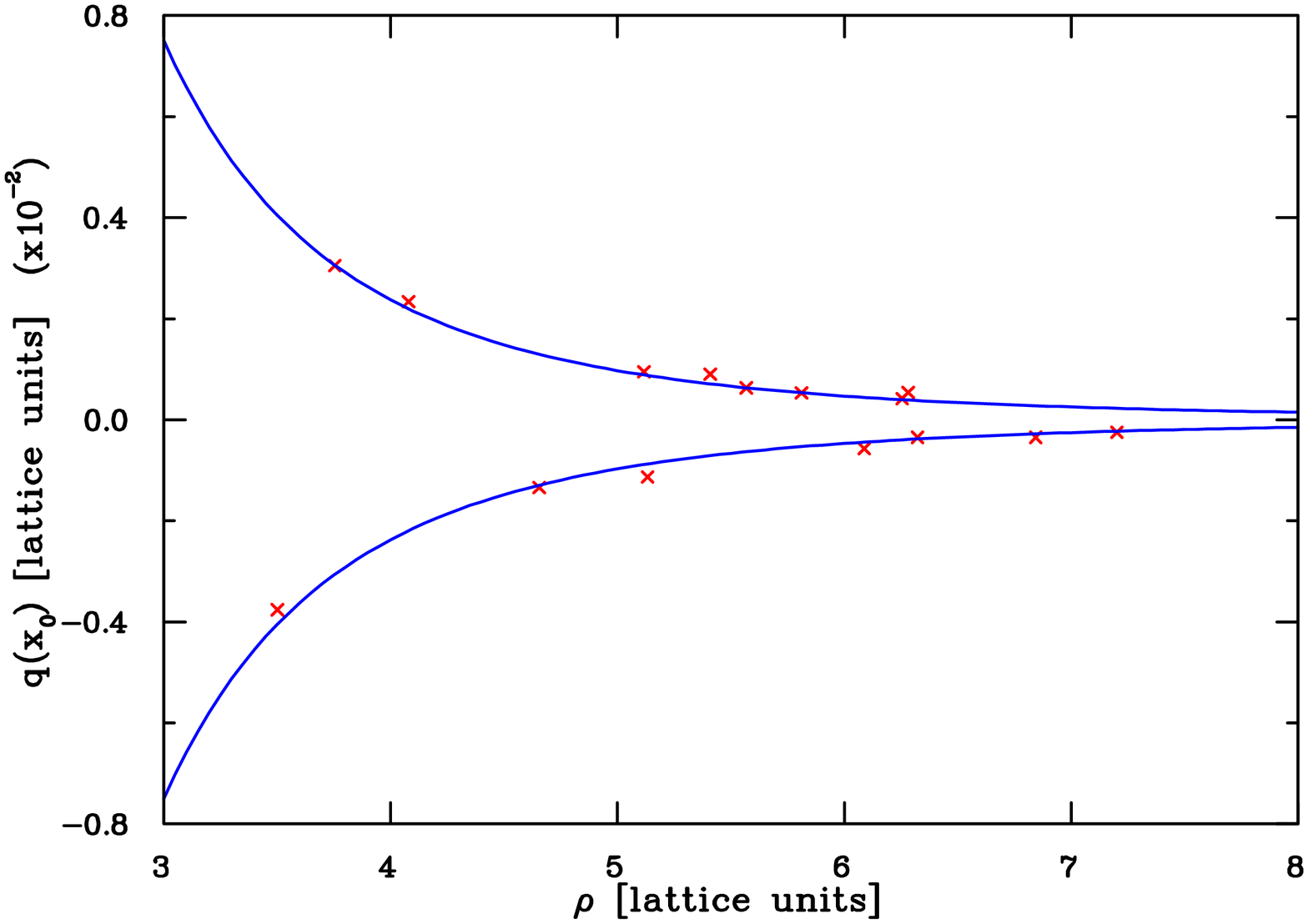}
    \caption{Three scatterplots showing the increasing instanton-like
     correlation between the gluonic topological charge density in the
     maxima $x_0$ of its modulus $|q_{sm}(x)|$ and the ``instanton radii'' 
     $\rho_{inst}$ (see text). The solid lines represent the 
     (anti)instanton-like relation between the two cluster parameters. 
     The upper plot shows a huge number of maxima after 5 smearing steps,
     without any relation between density and size. The middle and bottom 
     plots show a decreasing number of maxima and an inreasing accuracy of
     the instanton-like relation after 40 and 200 smearing steps, 
     respectively.}
    \label{fig:instantonlike}
  \end{center}
\end{figure*}

Using the matching of Eq.~(\ref{eq:matchqx}) we are able to directly
compare the overlap topological charge density with some level of UV
filtering to a given number of stout-link smearing sweeps. In an early
stadium of stout-link smearing the topological density does not yet
show classical, instanton-like features. 
What is meant by ``instanton-like features'' and how they gradually 
emerge from stout-link smearing is illustrated for a configuration 
with $Q=-1$ in Fig.~\ref{fig:instantonlike}. The solid lines represent
the relation
\begin{equation}
|q(x_0)| = \frac{6}{\pi^2 \rho_{inst}^4} \; ,
\end{equation}
typical for the (anti)instanton solution, between the gluonic topological 
charge density in the maxima $x_0$ of the modulus of the density $|q_{sm}(x)|$ 
and the ``instanton radii'' $\rho_{inst}$ obtained from a fit of the curvature 
of the action density in the points neighbouring $x_0$.
\begin{figure*}[h]
  \begin{center}
    \begin{tabular}{cc}
      \includegraphics[width=0.50\textwidth,angle=0]{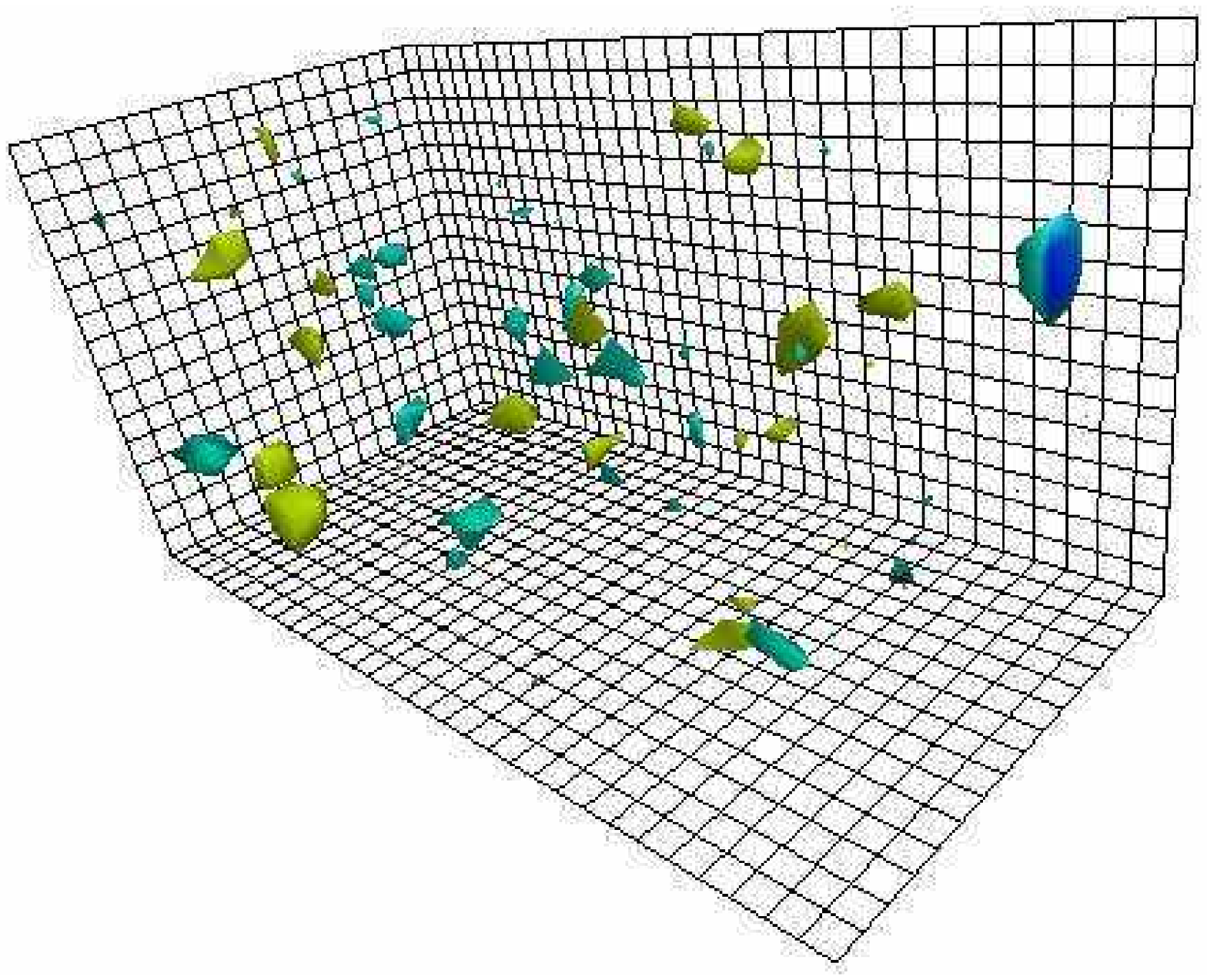} &
      \hspace{0.5cm}
      \includegraphics[width=0.50\textwidth,angle=0]{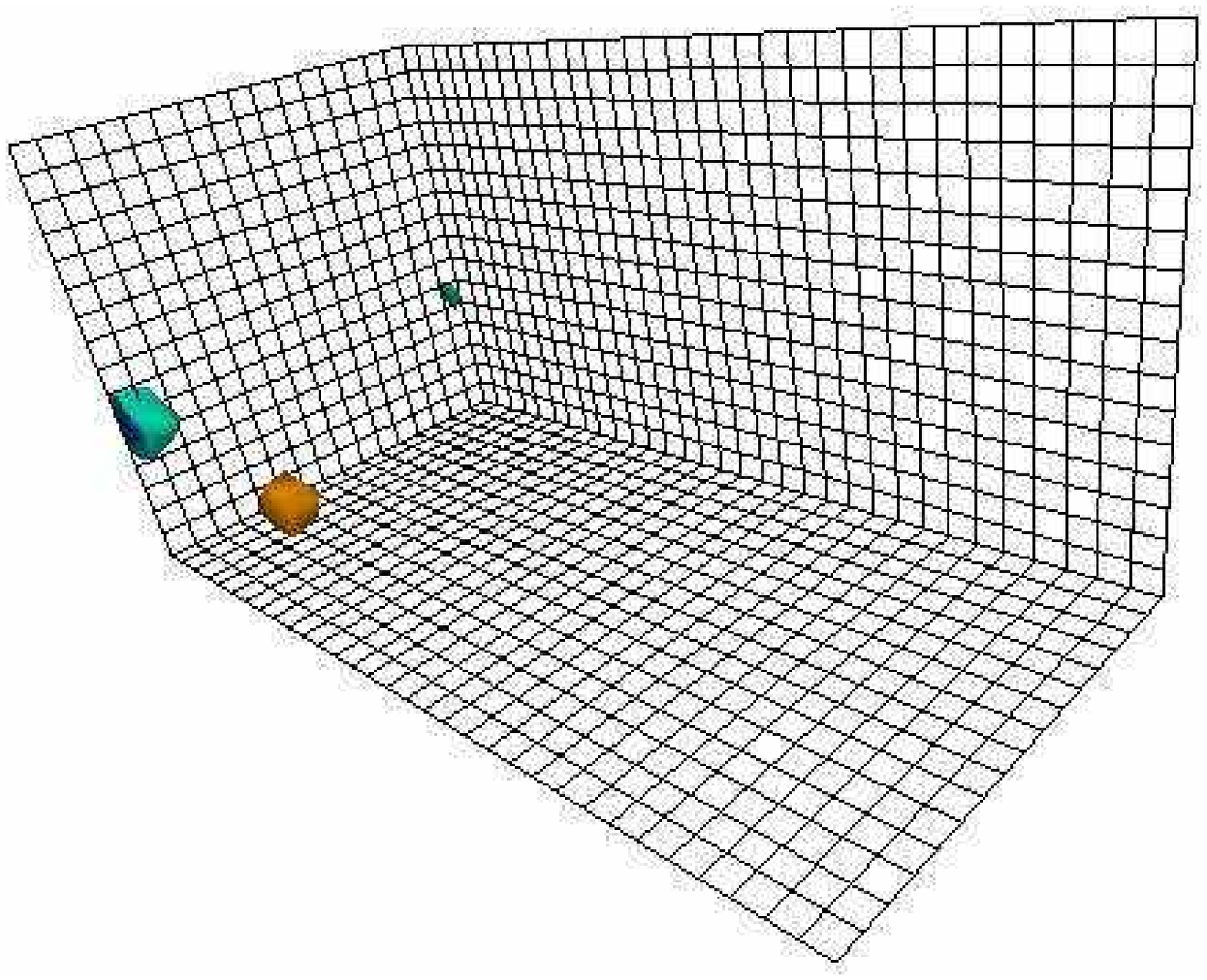} \\
    \end{tabular}
    \caption{The gluonic topological charge density very close to the
      maxima in some timeslice of the same configuration as analyzed
      in Fig.~\ref{fig:instantonlike} after 5 (left) and 40 stout-link
      smearing iterations (right). In color online: negative density
      blue/green, positive density red/yellow. In grey-scale:
      negative density dark, positive density light.}
  \label{fig:shapes}
  \end{center}
\end{figure*}
\begin{figure*}[h]
  \begin{center}
    \begin{tabular}{cc}
      \includegraphics[width=0.50\textwidth,angle=0]{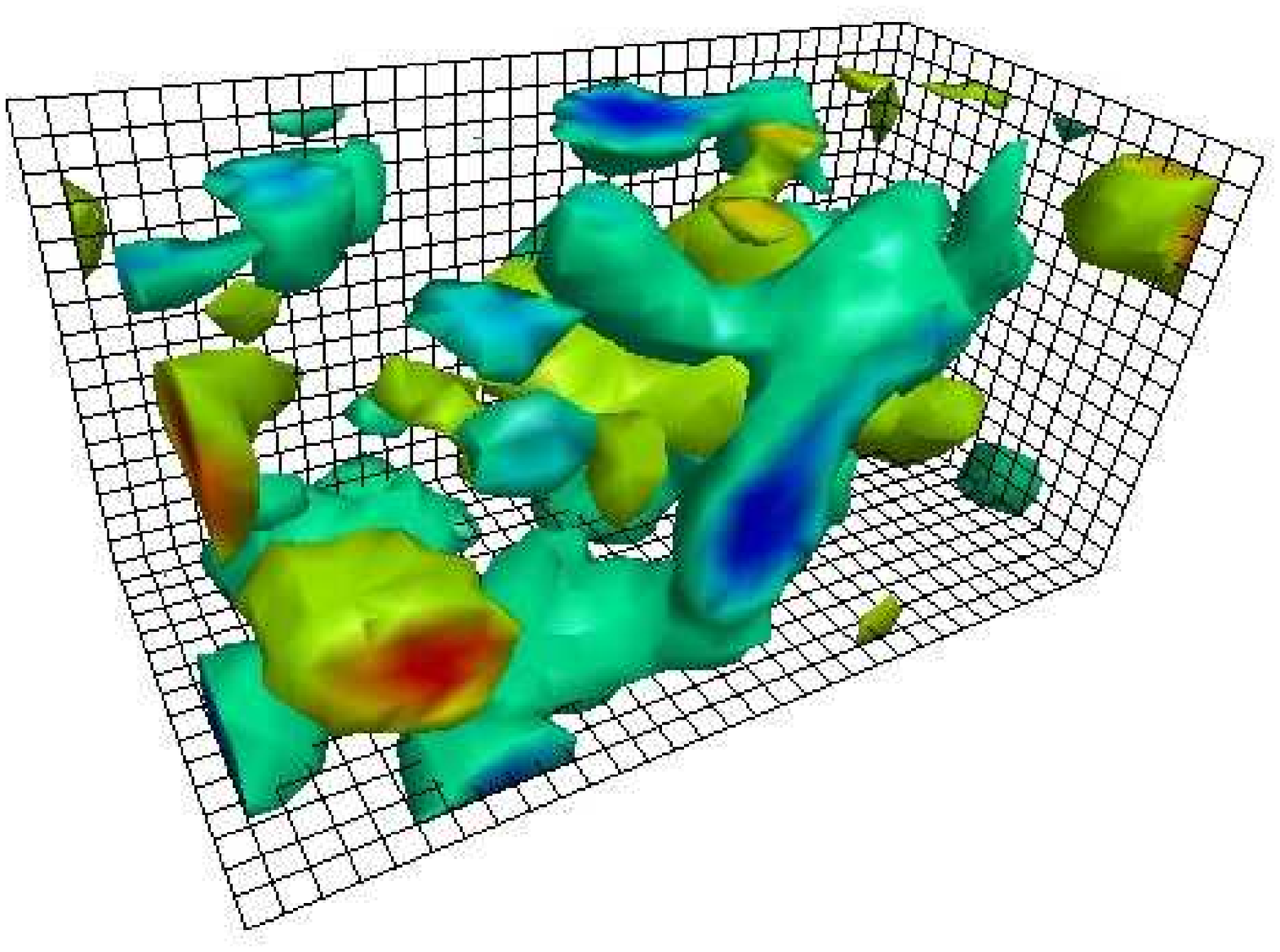} &
      \hspace{0.5cm}
      \includegraphics[width=0.50\textwidth,angle=0]{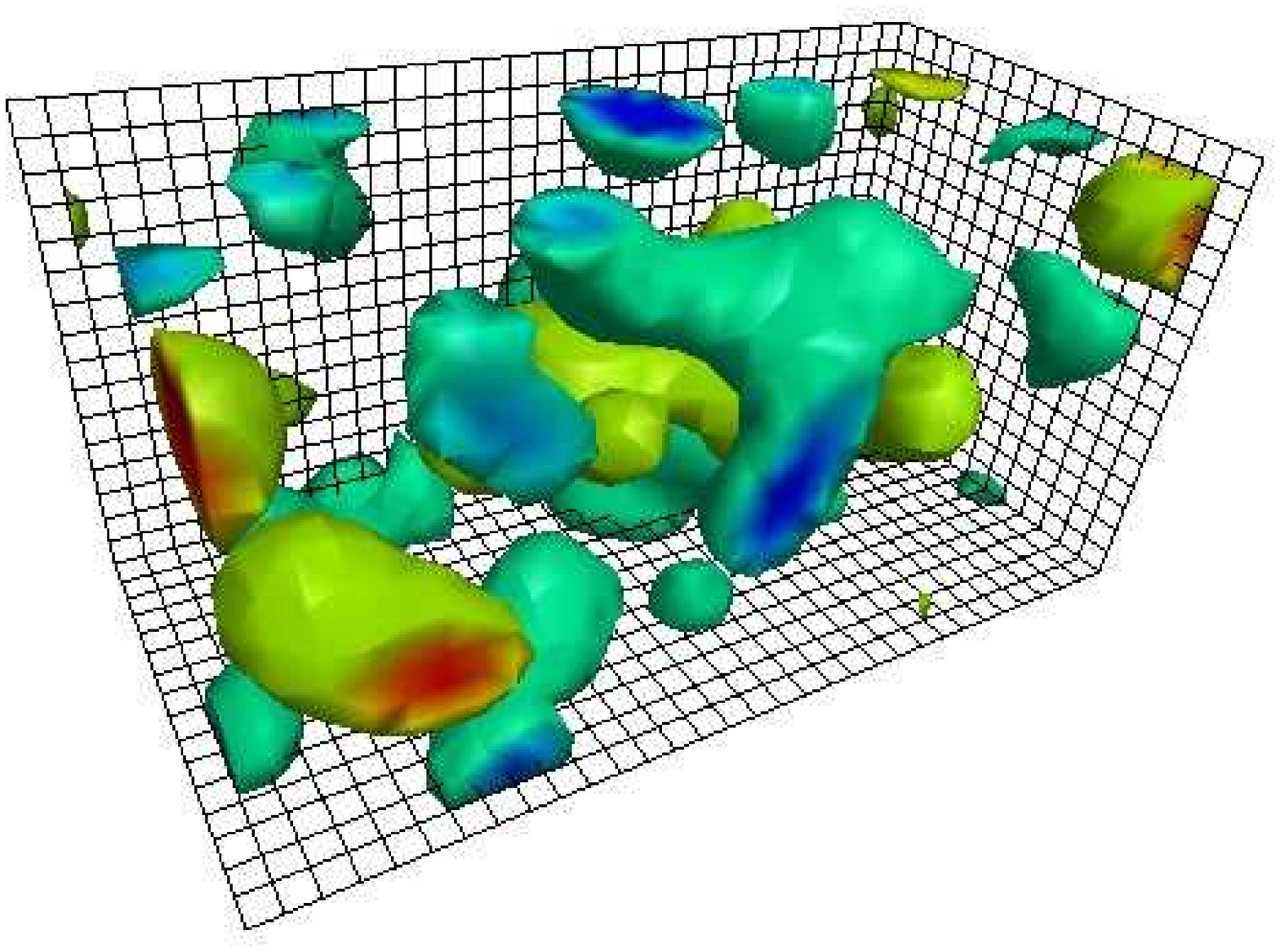} \\
    \end{tabular}
    \caption{The fermionic topological charge density of a $Q = 0$
      configuration with $\lambda_{cut}=634 \mathrm{~MeV}$ (left)
      compared with 48 sweeps of over-improved stout-link smearing
      (right). In color online: negative density blue/green,
      positive density red/yellow. In grey-scale: negative density
      dark, positive density light.}
  \label{fig:qx.634MeV}
  \end{center}
\end{figure*}
\begin{figure*}[h]
  \begin{center}
    \begin{tabular}{cc}
      \includegraphics[width=0.50\textwidth,angle=0]{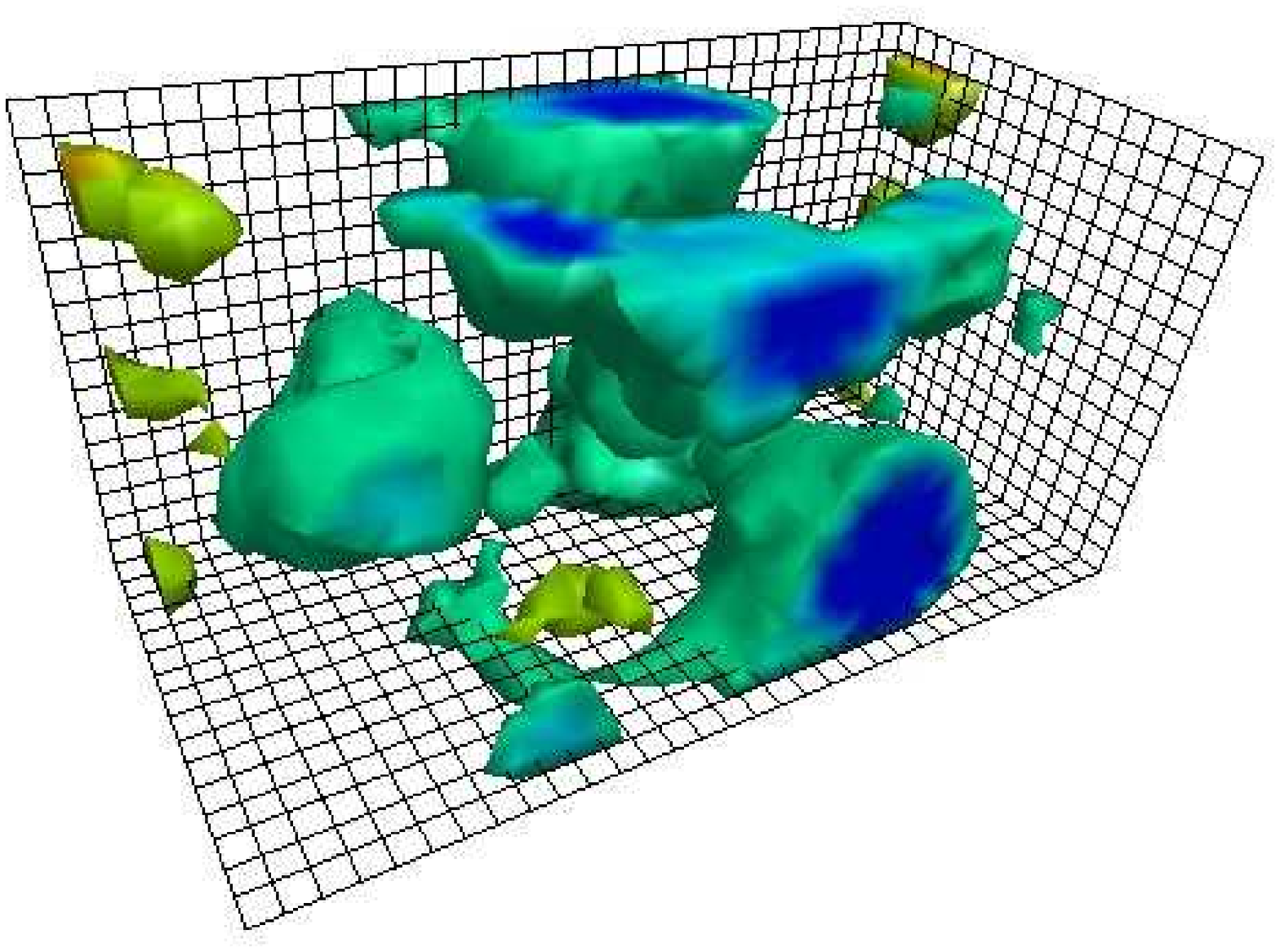} &
      \hspace{0.5cm}
      \includegraphics[width=0.50\textwidth,angle=0]{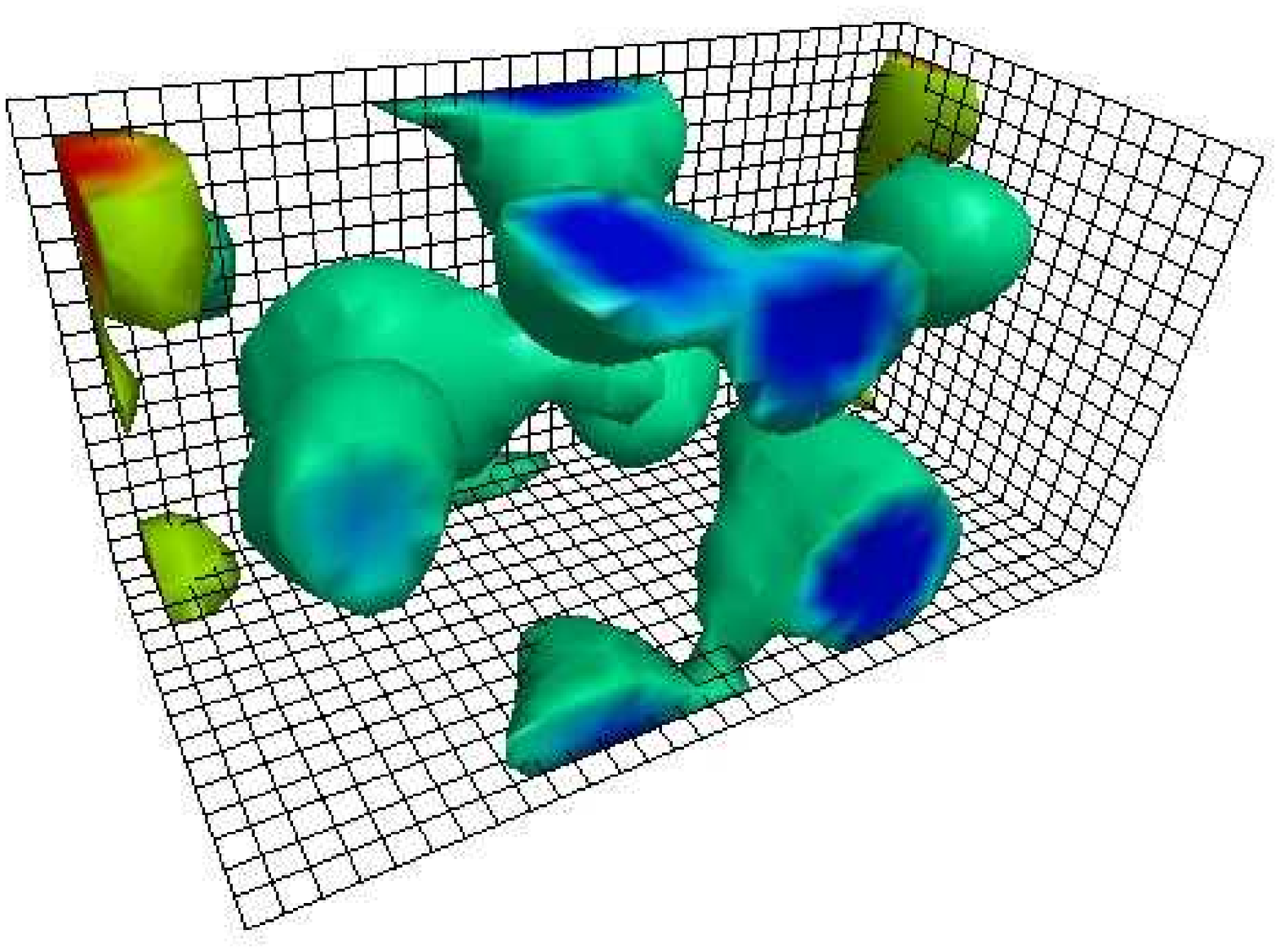} \\
    \end{tabular}
    \caption{The fermionic topological charge density of a $Q = -8$
      configuration with $\lambda_{cut}=400 \mathrm{~MeV}$ (left)
      compared with 103 sweeps of over-improved stout-link smearing
      (right). In color online: negative density blue/green,
      positive density red/yellow. In grey-scale: negative density
      dark, positive density light.}
  \label{fig:qx.400MeV}
  \end{center}
\end{figure*}
\begin{figure*}[h]
  \begin{center}
    \begin{tabular}{cc}
      \includegraphics[width=0.50\textwidth,angle=0]{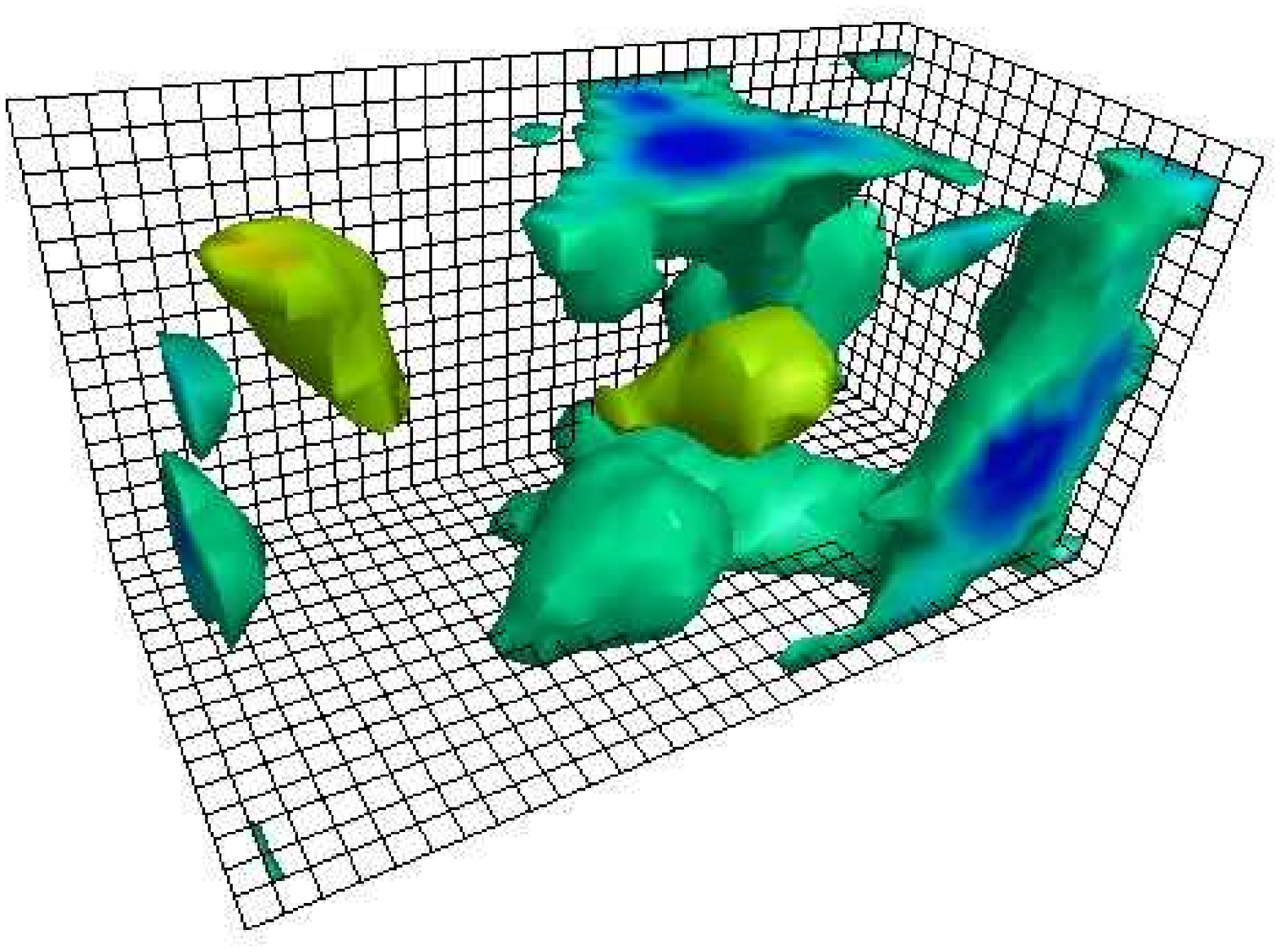} &
      \hspace{0.5cm}
      \includegraphics[width=0.50\textwidth,angle=0]{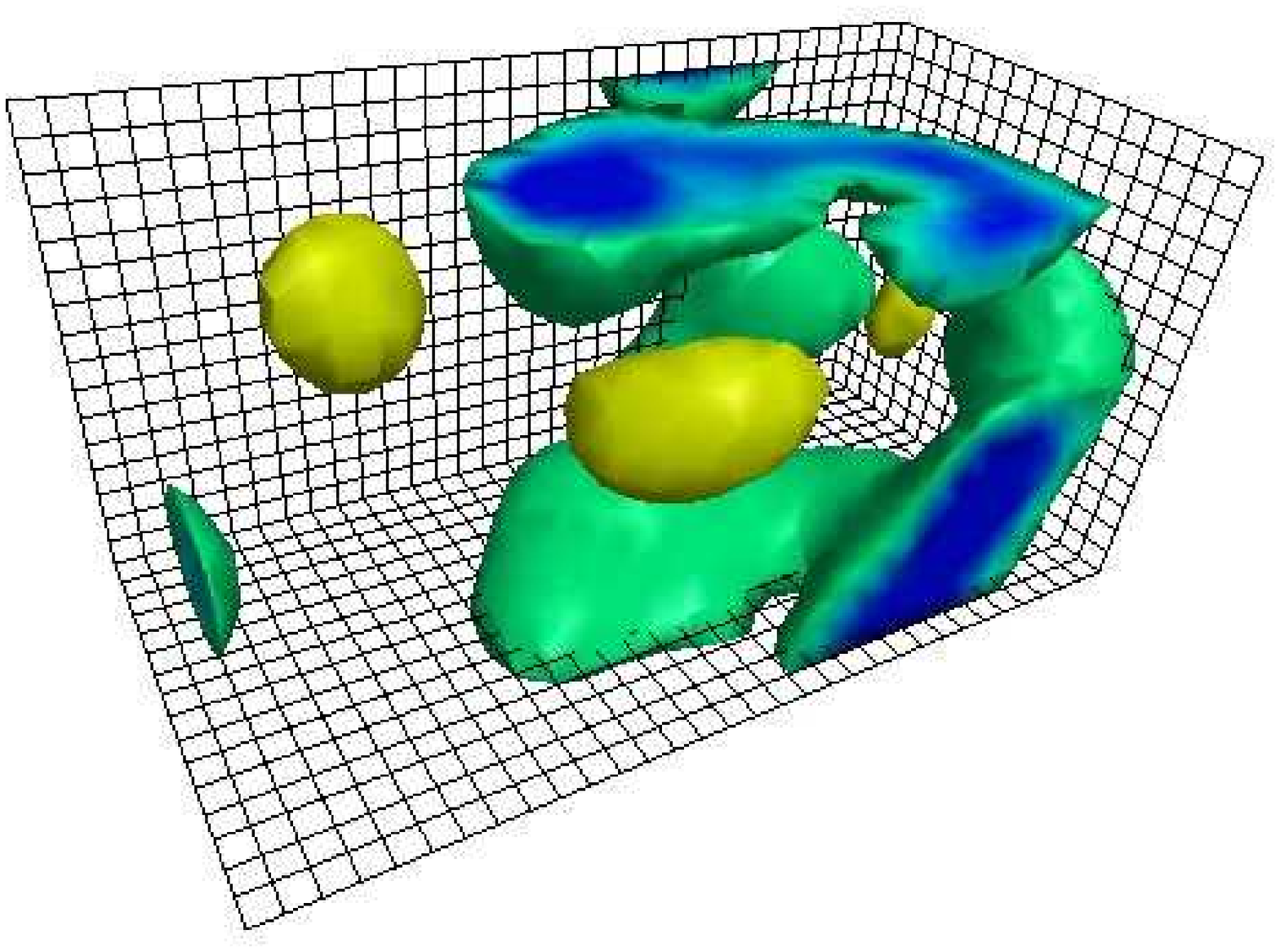} \\
    \end{tabular}
    \caption{The fermionic topological charge density of a $Q = -7$
      configuration with $\lambda_{cut}=200 \mathrm{~MeV}$ (left)
      compared with 192 sweeps of over-improved stout-link smearing
      (right). In color online: negative density blue/green,
      positive density red/yellow. In grey-scale: negative density
      dark, positive density light.}
  \label{fig:qx.200MeV}
  \end{center}
\end{figure*}
The upper plot has to be considered with a grain of salt because a closer
look at the maxima reveals a substantial lack of isotropy of the peaks 
of topological density which is however implicitely assumed in 
the fits of $\rho_{inst}$. 
The marked difference in this respect between 5 and 40 smearing steps is 
visible in Fig.~\ref{fig:shapes} showing a certain timeslice of the same 
configuration as in Fig.~\ref{fig:instantonlike}.

For a moderate amount of smearing and filtering, respectively, the two
topological density definitions are faithfully exhibiting the
outstanding clusterization of charge, provided the cut-off
$\lambda_{cut}$ and the number of stout-link smearing steps are
optimally tuned to each other.
This is exemplified by Fig.~\ref{fig:qx.634MeV} which shows the same
time-slice of a $Q=0$ configuration, on the left-hand side portrayed
by the overlap-fermionic topological density with $\lambda_{cut}=634
\mathrm{~MeV}$ and on the right-hand side by the gluonic topological
density after 48 stout-link smearing steps.

In Fig.~\ref{fig:qx.400MeV} we show the fermionic density of the $Q =
-8$ configuration, with $\lambda_{cut}=400\mathrm{~MeV}$ and the
gluonic density after 103 stout-link smearing steps. In this stadium
of smearing, the bias in favor of classical lumps is already visible.

Besides the similarity between the two methods, the tendency towards
classical lumps becomes obvious in Fig.~\ref{fig:qx.200MeV}.  This
figure shows the $Q = -7$ configuration with a cutoff of
$\lambda_{cut} = 200~\mathrm{MeV}$. At this level of UV filtering the
best match is provided by 193 sweeps of over-improved stout-link
smearing. In this stadium of smearing, in particular because of the
over-improved action built into the smearing, the minority 
positive charge has
become stabilized compared with what the fermionic view shows.

\subsection{Zero modes and lowest non-zeromodes and the instanton
  content after long smearing}
\label{subsec:cluster2}

For long smearing with essentially 
more than 100 smearing steps, the similarity of
the topological densities slowly becomes less perfect. Still, the
position of the gluonic topological lumps is not completely arbitrary
from the point of view of the original configuration. The zero modes
(if $Q\ne0$) and the lowest non-zero modes contain a high predictive
power over where these lumps will appear.

\begin{figure*}[h]
  \begin{center}
    \begin{tabular}{cc}
      \includegraphics[width=0.50\textwidth,angle=0]{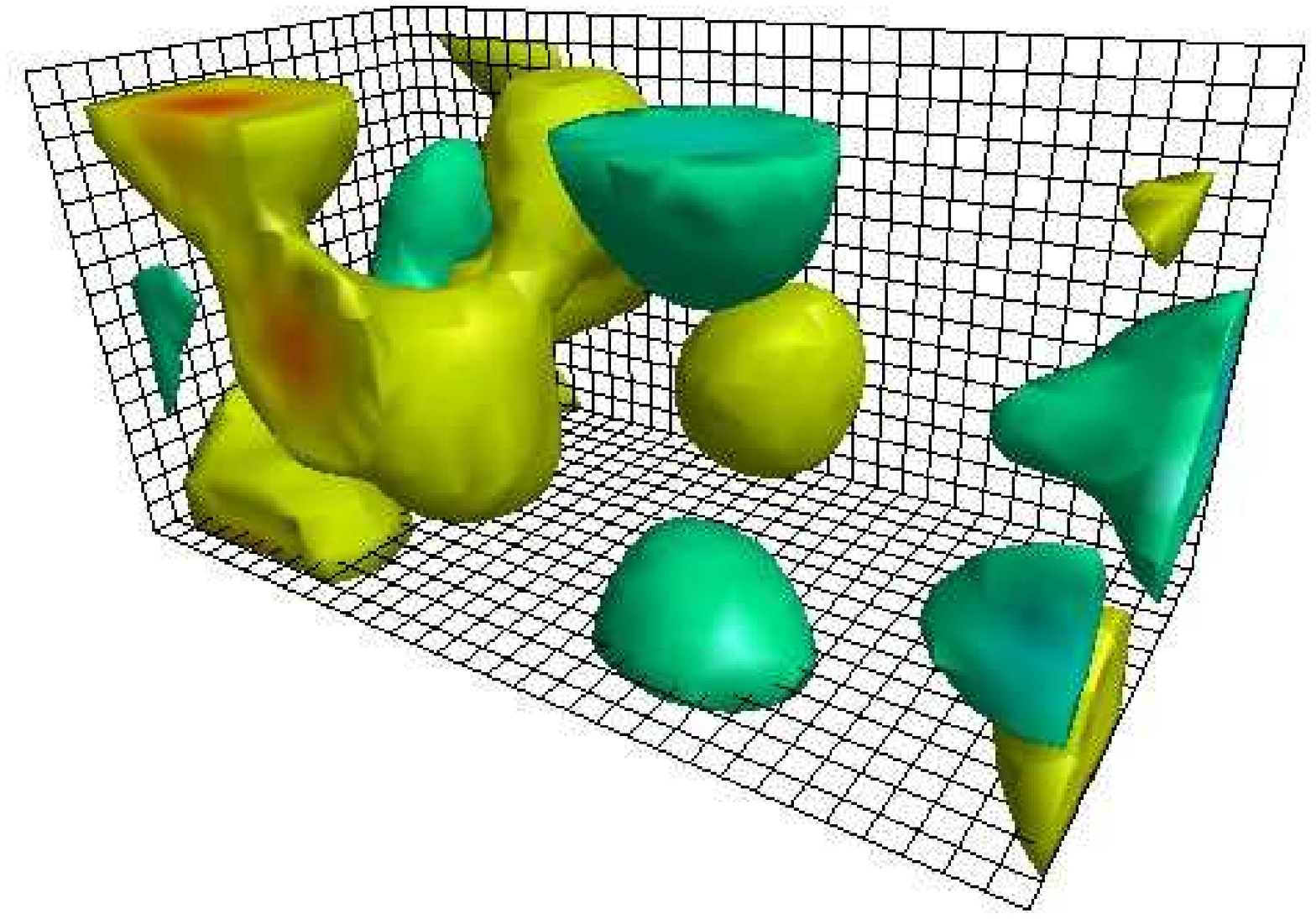} &
      \hspace{0.5cm}
      \includegraphics[width=0.50\textwidth,angle=0]{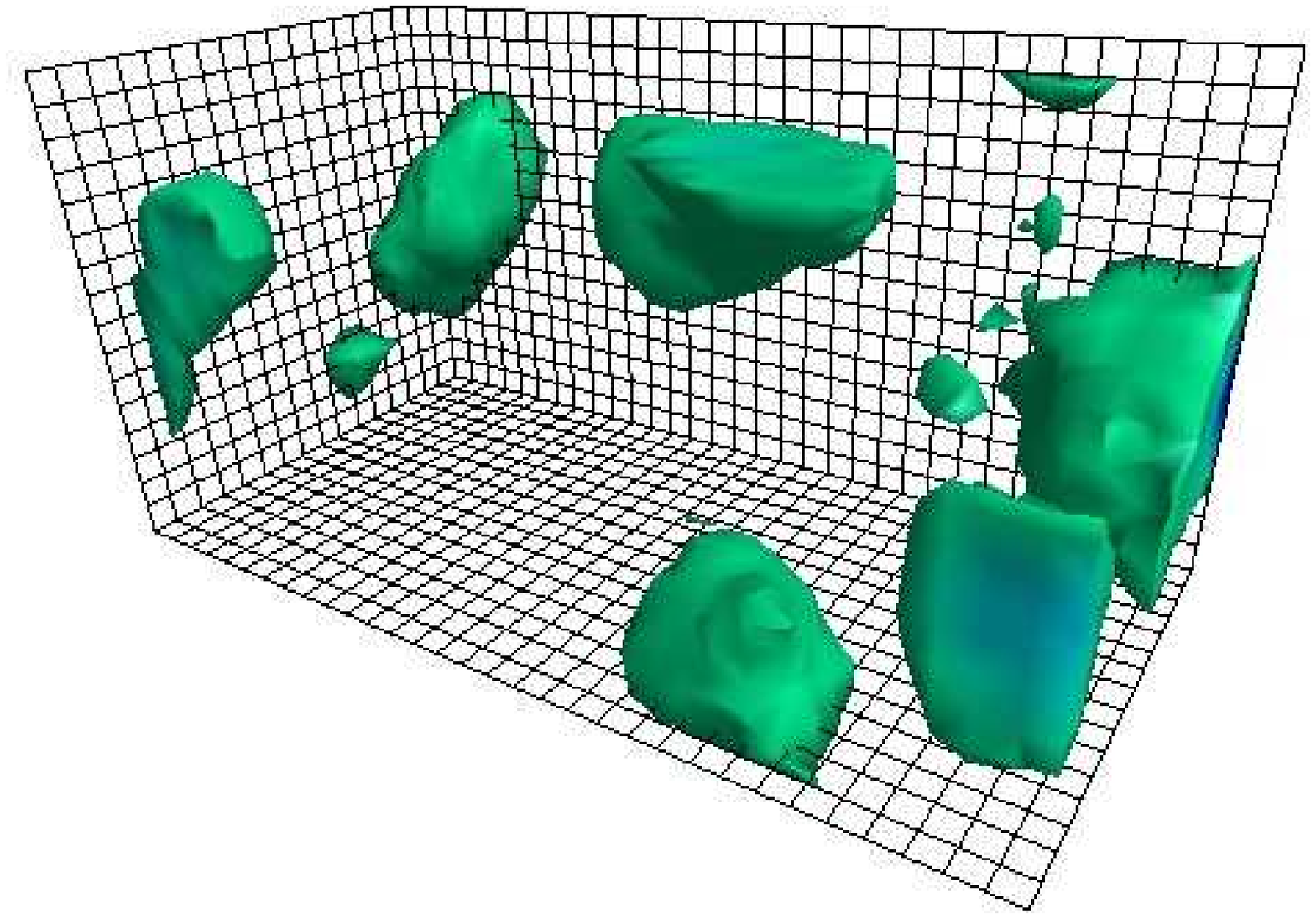} \\ 
    \end{tabular}
    \caption{The gluonic topological charge density after 200 smearing
      sweeps (left) and the scalar density of the zero mode (right)
      for a $Q=-1$ configuration. One can see how the zero mode
      extends over only regions of negative charge. All extended
      sign-coherent objects seen are good local approximations to
      instantons or anti-instantons in the centre. In color online:
      negative density is blue/green, positive density is red/yellow
      (left), and scalar density is blue/green (right). In grey-scale:
      negative density dark, positive density light (left).}
  \label{fig:zm.0123-rho-0}
  \end{center}
\end{figure*}
\begin{figure*}[h]
  \begin{center}
    \begin{tabular}{cc}
      \includegraphics[width=0.50\textwidth,angle=0]{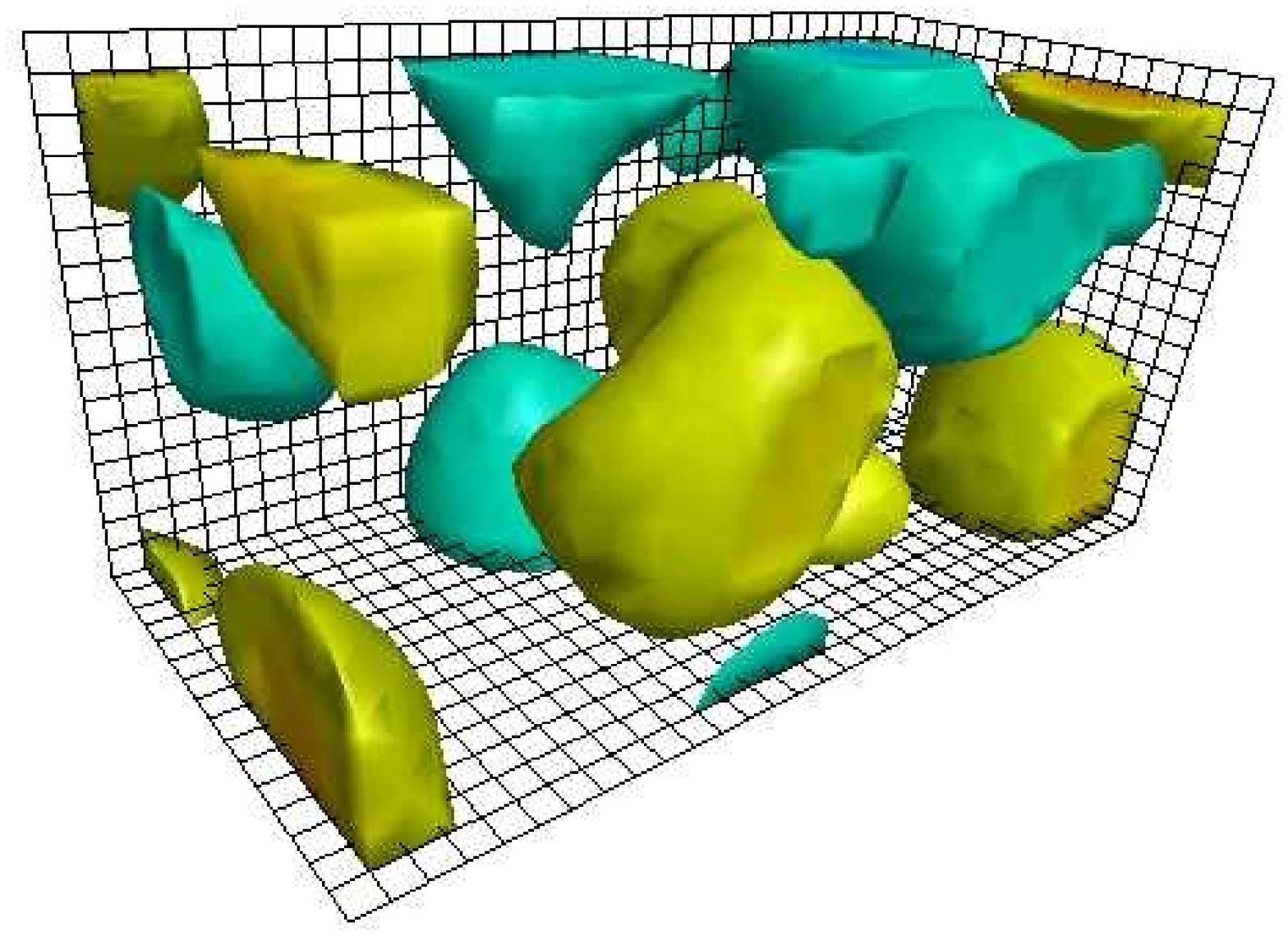} &
      \hspace{0.5cm}
      \includegraphics[width=0.50\textwidth,angle=0]{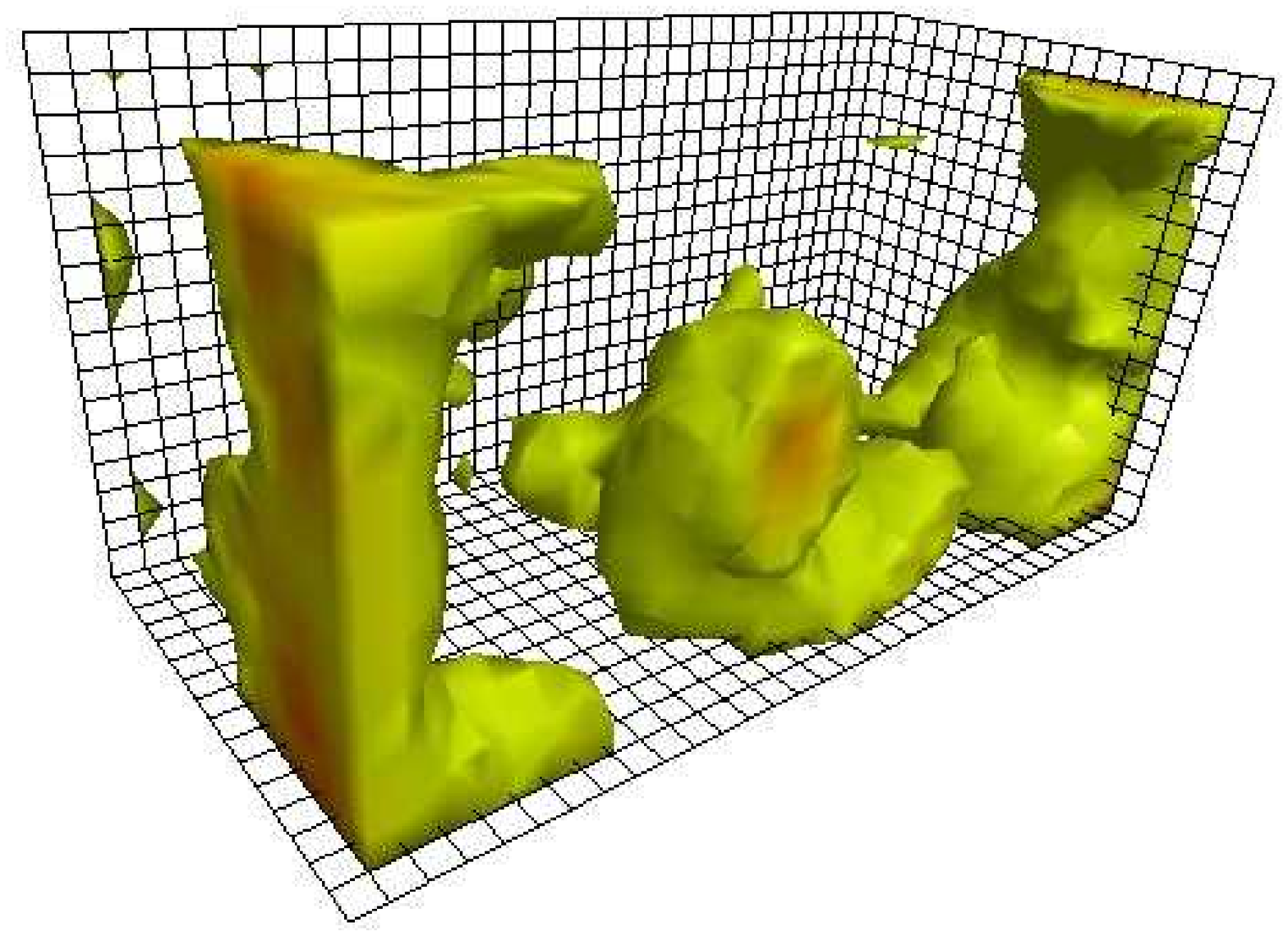} \\ 
    \end{tabular}
    \caption{The gluonic topological charge density after 200 smearing
      sweeps (left) and the scalar density of the zero mode (right)
      for a $Q=1$ configuration. In this case the high density regions
      of the zero mode are centred on lumps of positive charge. Again
      all objects are good approximations to classical instantons. In
      color online: negative density is blue/green, positive density
      is red/yellow (left), and scalar density is red/yellow
      (right). In grey-scale: negative density dark, positive density
      light (left).}
  \label{fig:zm.0130-rho-0}
  \end{center}
\end{figure*}
\begin{figure*}[!ht]
  \begin{center}
    \begin{tabular}{cc}
      \includegraphics[width=0.50\textwidth,angle=0]{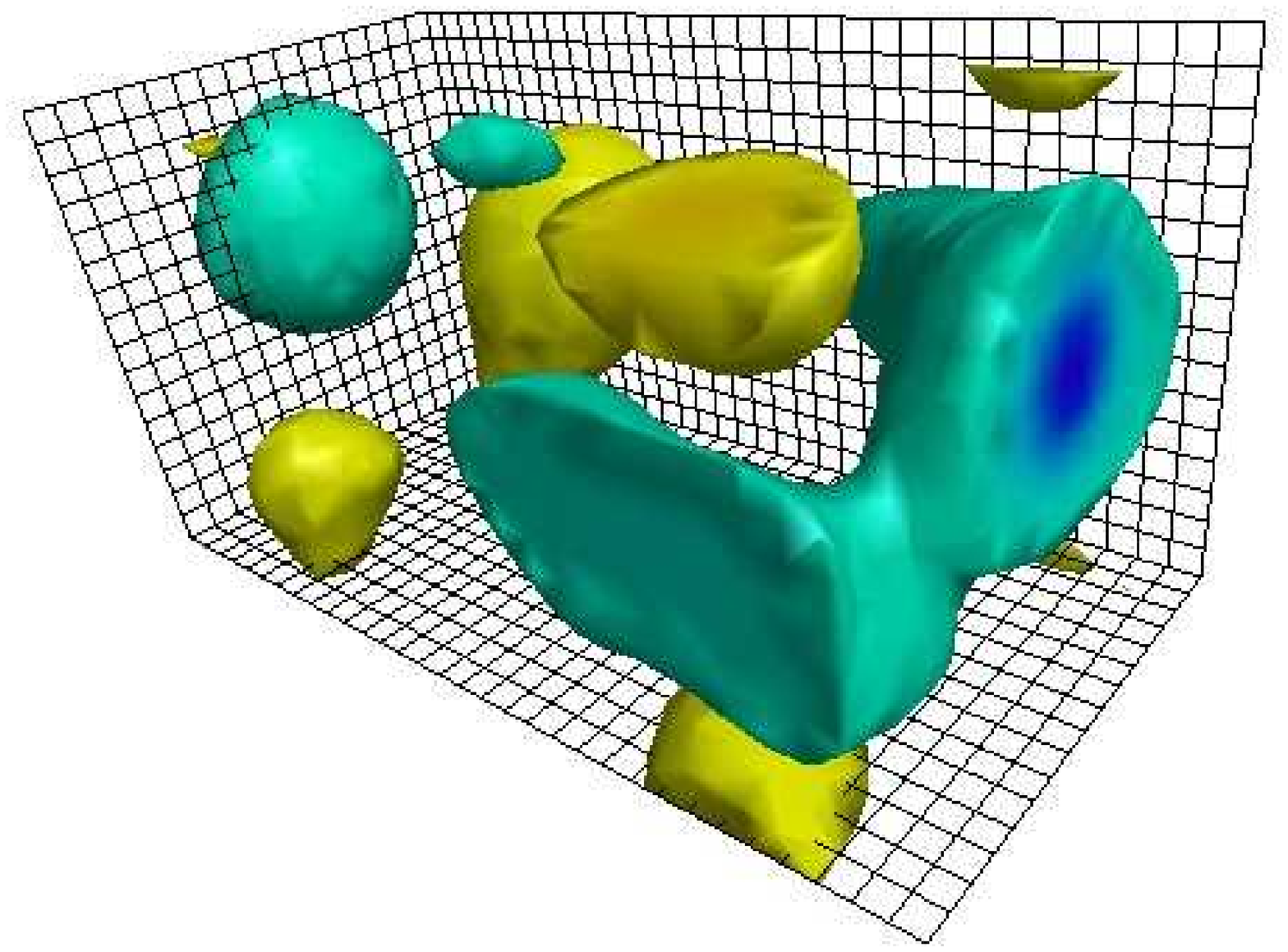} &
      \hspace{0.5cm}
      \includegraphics[width=0.50\textwidth,angle=0]{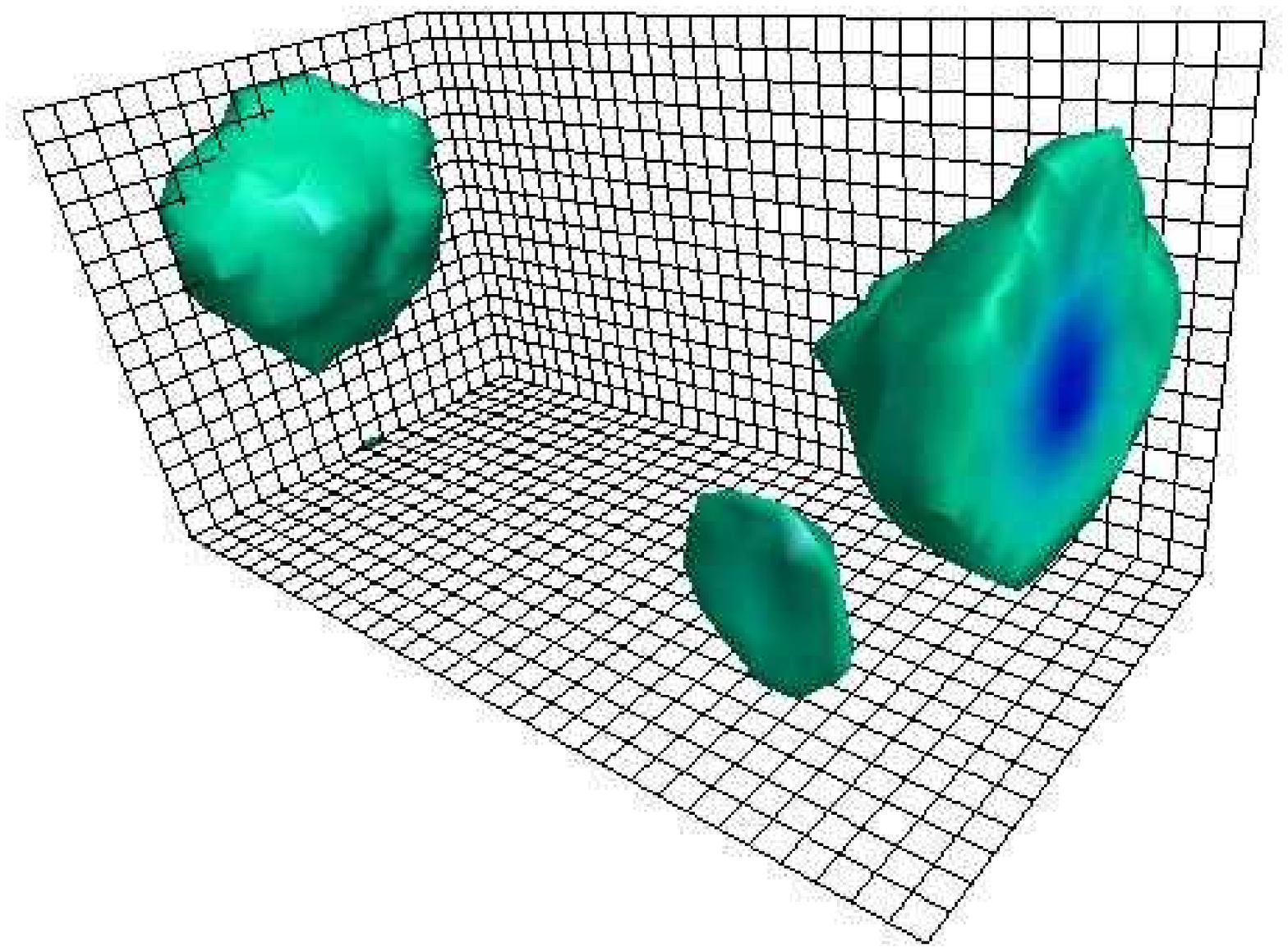} \\ 
    \end{tabular}
    \caption{The gluonic topological charge density after 200 smearing
      sweeps (left) and the scalar density of the zero mode (right)
      for another $Q=-1$ configuration. The high density regions of
      the zero mode are again centred on lumps of negative charge. All
      topological objects are good approximations to classical
      instantons. In color online: negative density is blue/green,
      positive density is red/yellow (left), and scalar density is
      blue/green (right). In grey-scale: negative density dark,
      positive density light (left).}
  \label{fig:zm.0141-rho-0}
  \end{center}
\end{figure*}

For 200 smearing steps, this is illustrated in
Fig.~\ref{fig:zm.0123-rho-0} by a $Q=-1$ configuration. In the left
panel the gluonic topological density is shown, in the right panel the
scalar density of the zero mode. The zero mode covers three distinct
centers of topological charge of appropriate sign only in the selected
time-slice. 
Two more examples of total charge $Q=\pm1$ are shown in 
Figs.~\ref{fig:zm.0130-rho-0}
and~\ref{fig:zm.0141-rho-0}. One sees that the zero mode does not
always cover all regions of appropriate charge. In other words, the
gluonic version of the topological charge density for some clusters 
- even in a late stadium of smearing - may be built by non-zero modes.
This leads us to revise the naive expectations according to which each 
zero mode would be residing on one lump of excess topological charge, 
for which a cluster charge of $Q_{cl}=\pm1$ would be suggested by the 
instanton model.
In fact, in Ref.~\cite{Ilgenfritz:2007ua} it has been demonstrated that 
the zero modes typically are simultaneously carried by a moderate number
of centers below the peak density, before they start percolating throughout 
the whole lattice at an even lower level of the scalar density. Here we 
additionally learn that all theses centers are marked by lumps of topological 
charge of appropriate sign, however not necessarily all lumps are covered.

\begin{figure*}[!h]
  \begin{center}
      \includegraphics[width=0.54\textwidth,angle=0]{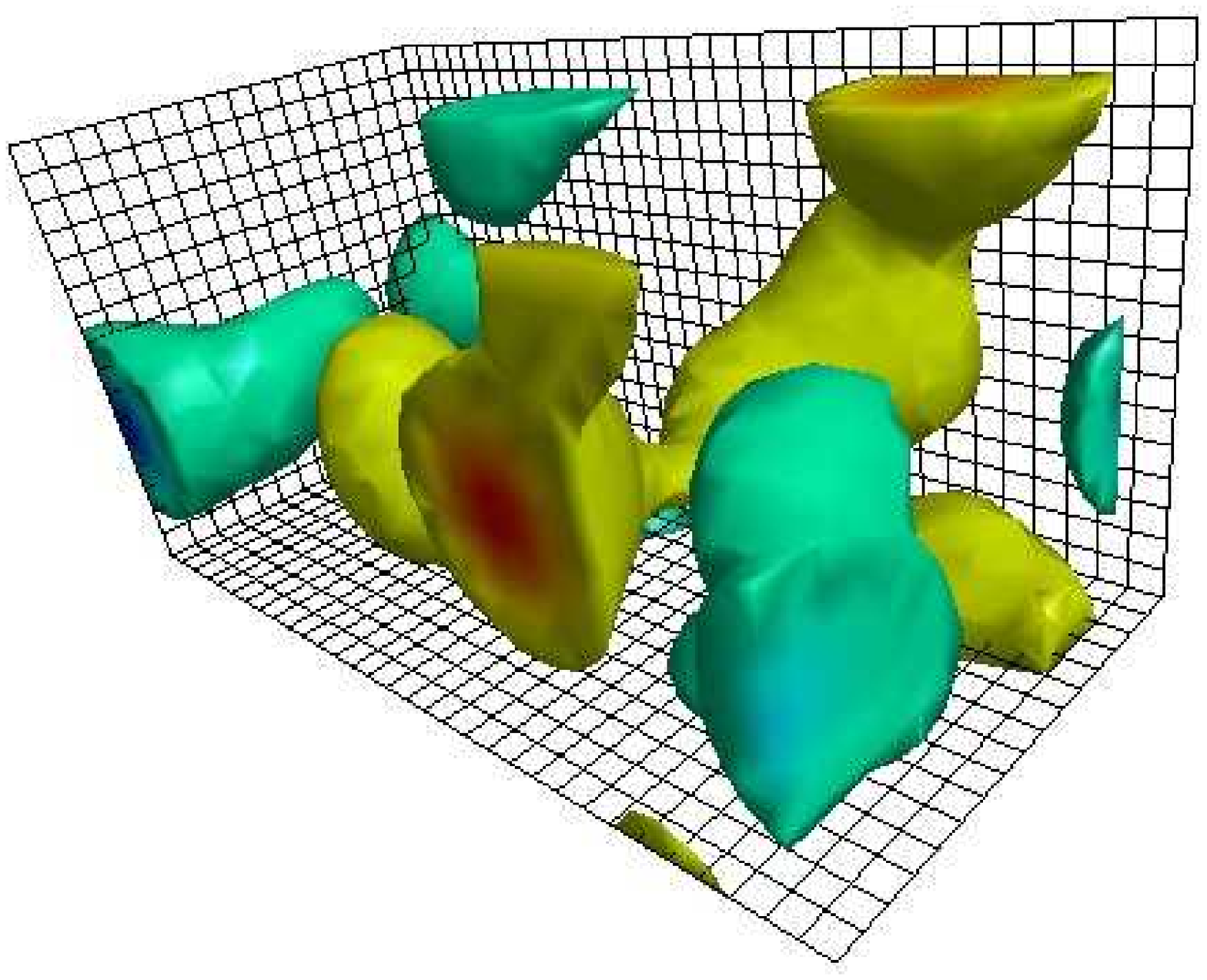} \\
      \hspace{0.5cm}
      \includegraphics[width=0.54\textwidth,angle=0]{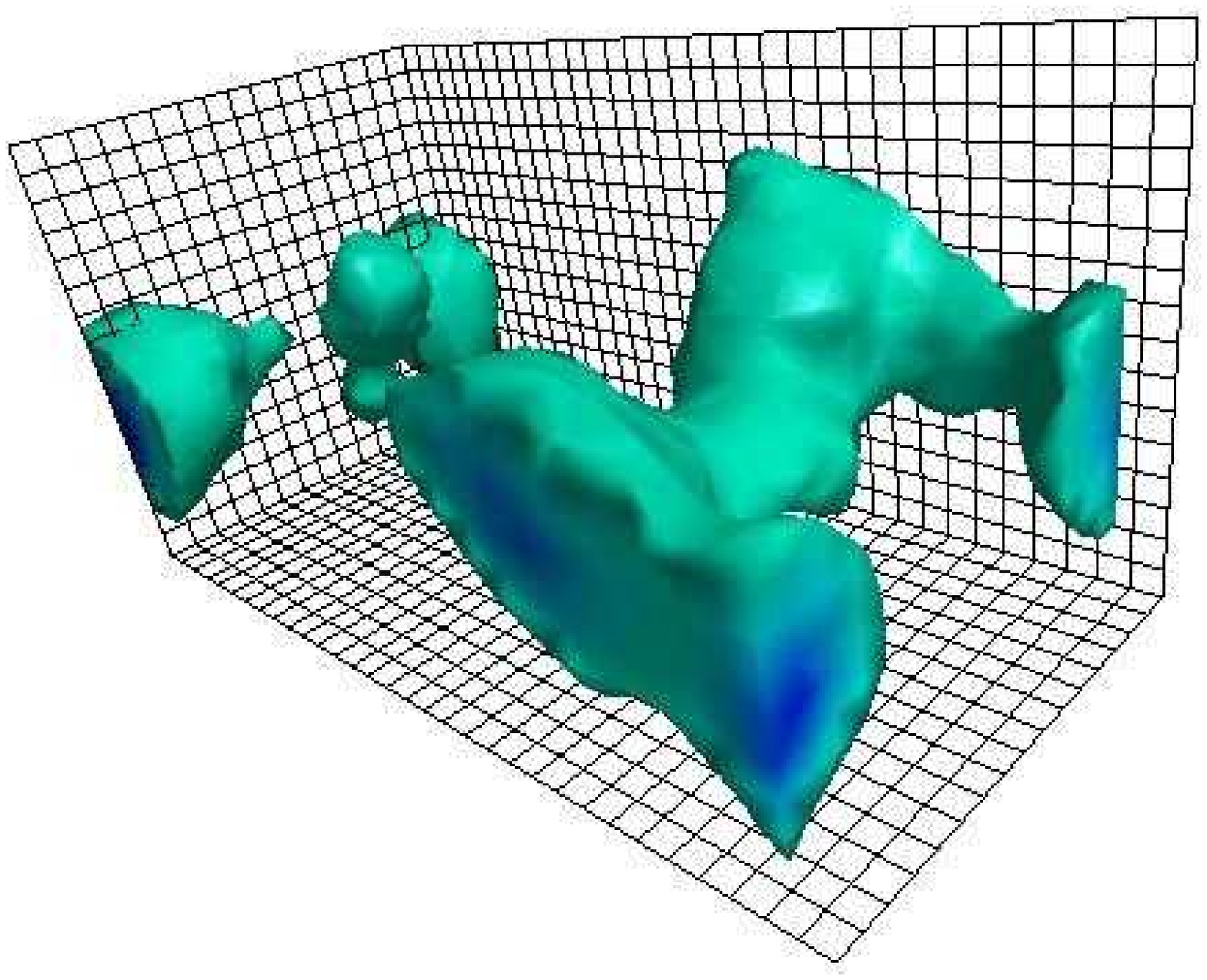} \\ 
      \hspace{0.5cm}
      \includegraphics[width=0.54\textwidth,angle=0]{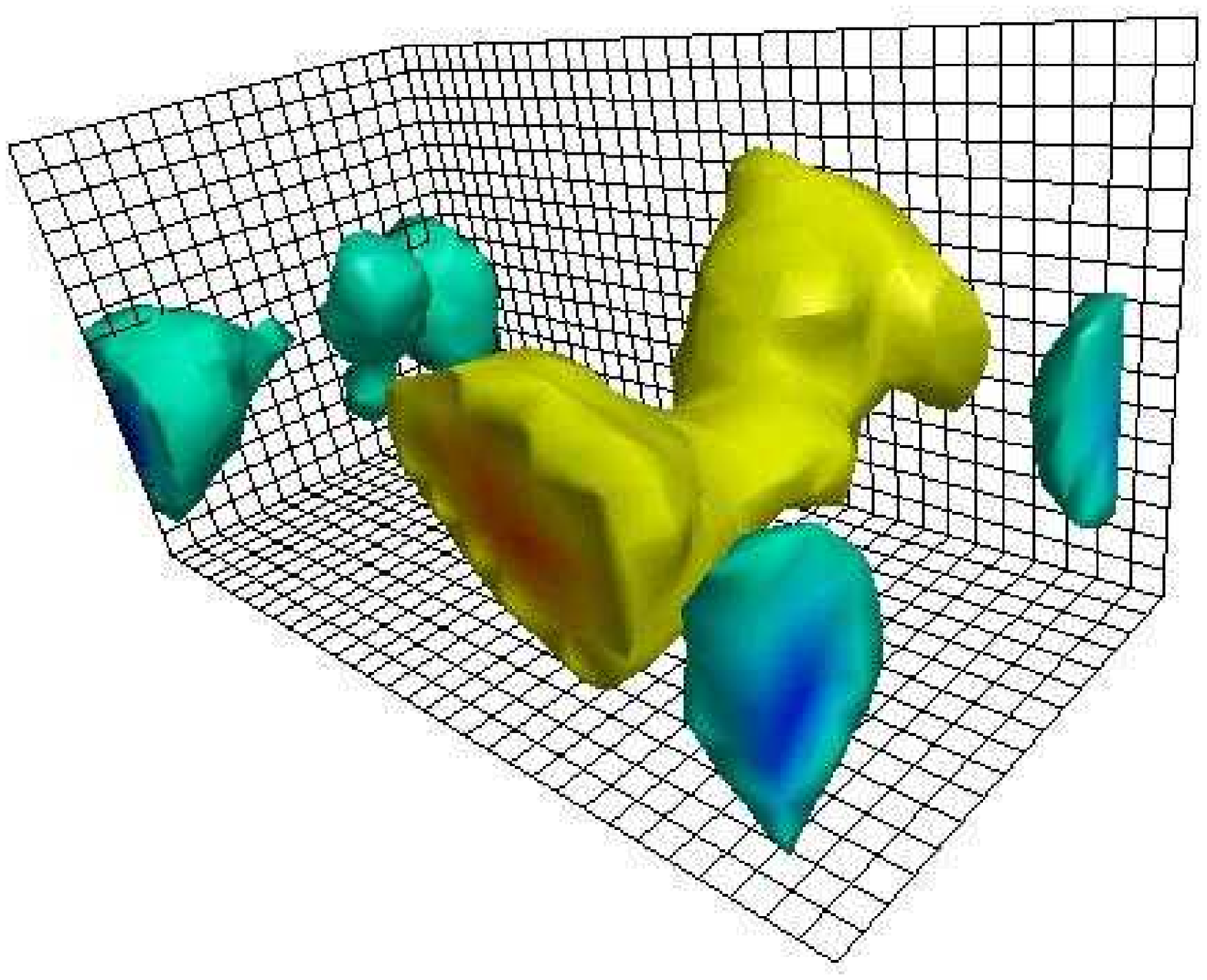}
      \caption{The scalar (middle) and pseudoscalar (bottom) density
        of the first non-zero mode for the $Q=-1$ configuration shown
        previously in Fig.~\ref{fig:zm.0123-rho-0}, along with the
        gluonic topological charge density after 200 sweeps of
        smearing (top). One sees how the scalar density extends over
        objects of differing charge, but that the regions of alternate
        charge are realized by the local chirality of the pseudoscalar
        density. In color online: negative density is blue/green,
        positive density is red/yellow (top), positive chirality is
        blue/green and negative chirality is red/yellow (bottom). In
        grey-scale: negative density dark, positive density light
        (top), and positive chirality dark, negative chirality light
        (bottom).}
    \label{fig:zm.0123-rho-1}
  \end{center}
\end{figure*}

For these same three configurations we now consider the distribution
of the lowest non-zero modes. Starting with the first $Q=-1$
configuration, we present the scalar and pseudoscalar densities of the
first non-zero mode in Fig.~\ref{fig:zm.0123-rho-1}.
One sees that the pseudoscalar density, according to its local
chirality, highlights certain parts of the topological lumps with
appropriate sign of charge, and leaves others (for other low-lying
modes).

\begin{figure*}[!h]
  \begin{center}
      \includegraphics[width=0.54\textwidth,angle=0]{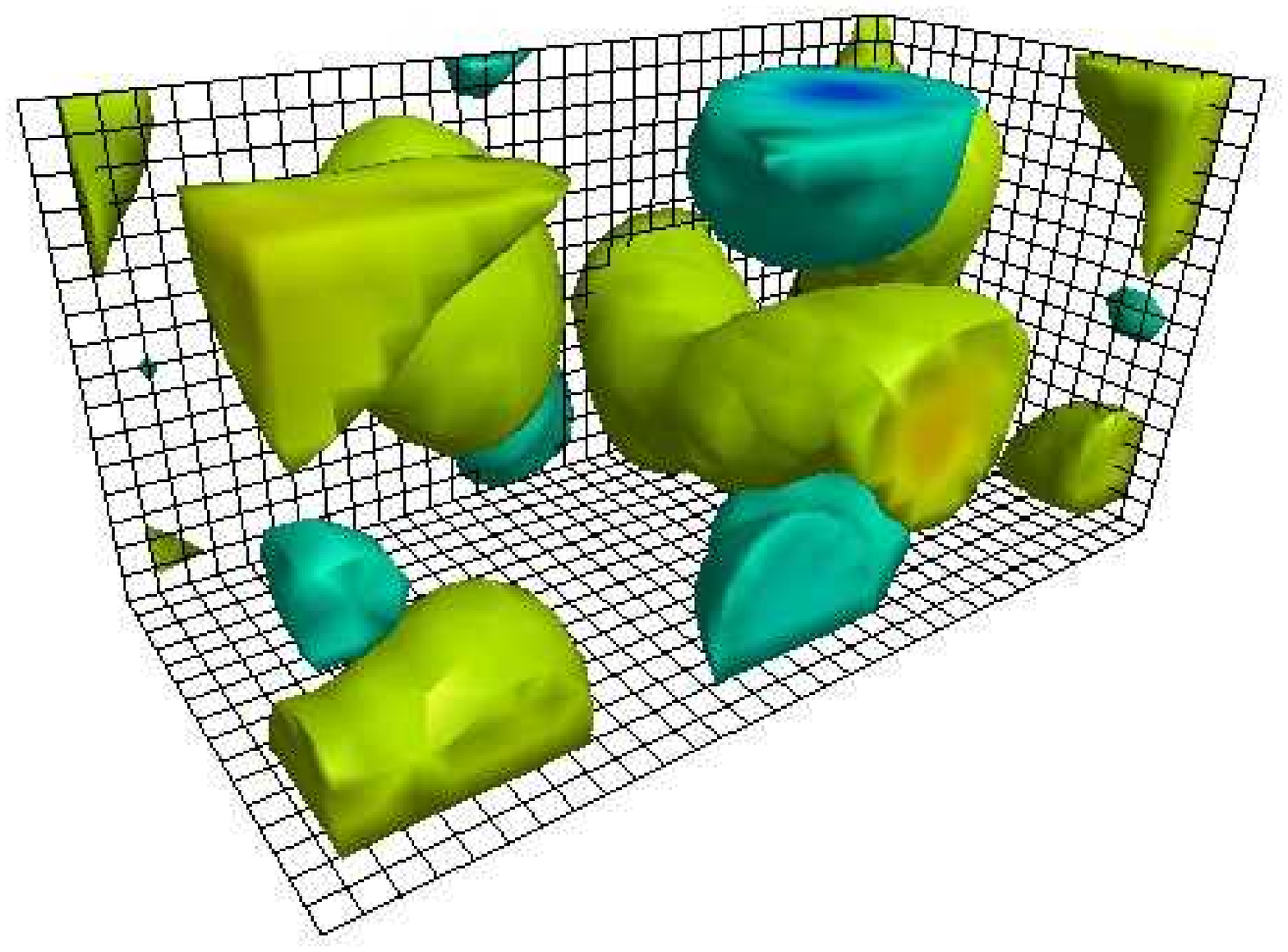} \\
      \hspace{0.5cm}
      \includegraphics[width=0.54\textwidth,angle=0]{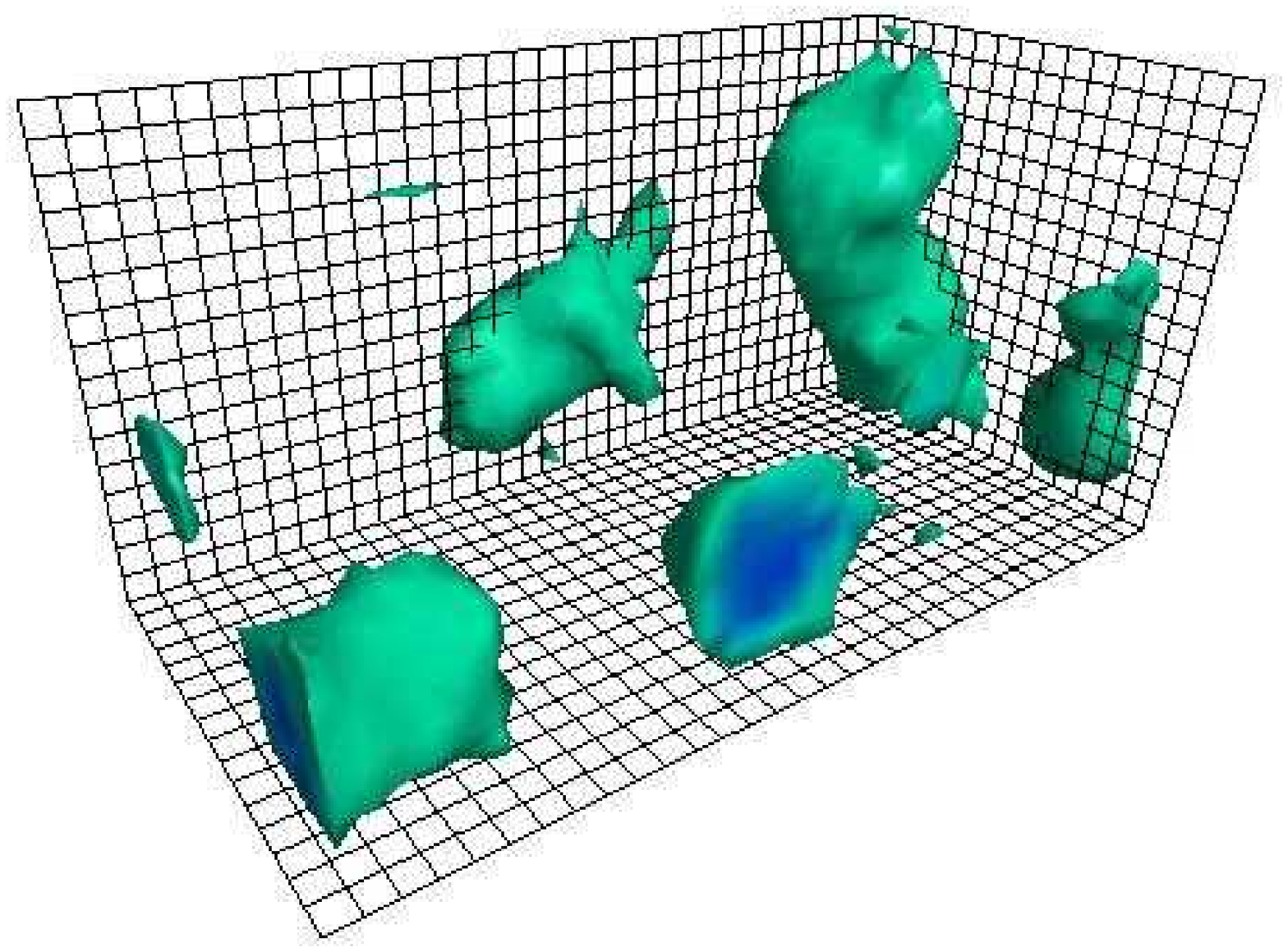} \\ 
      \hspace{0.5cm}
      \includegraphics[width=0.54\textwidth,angle=0]{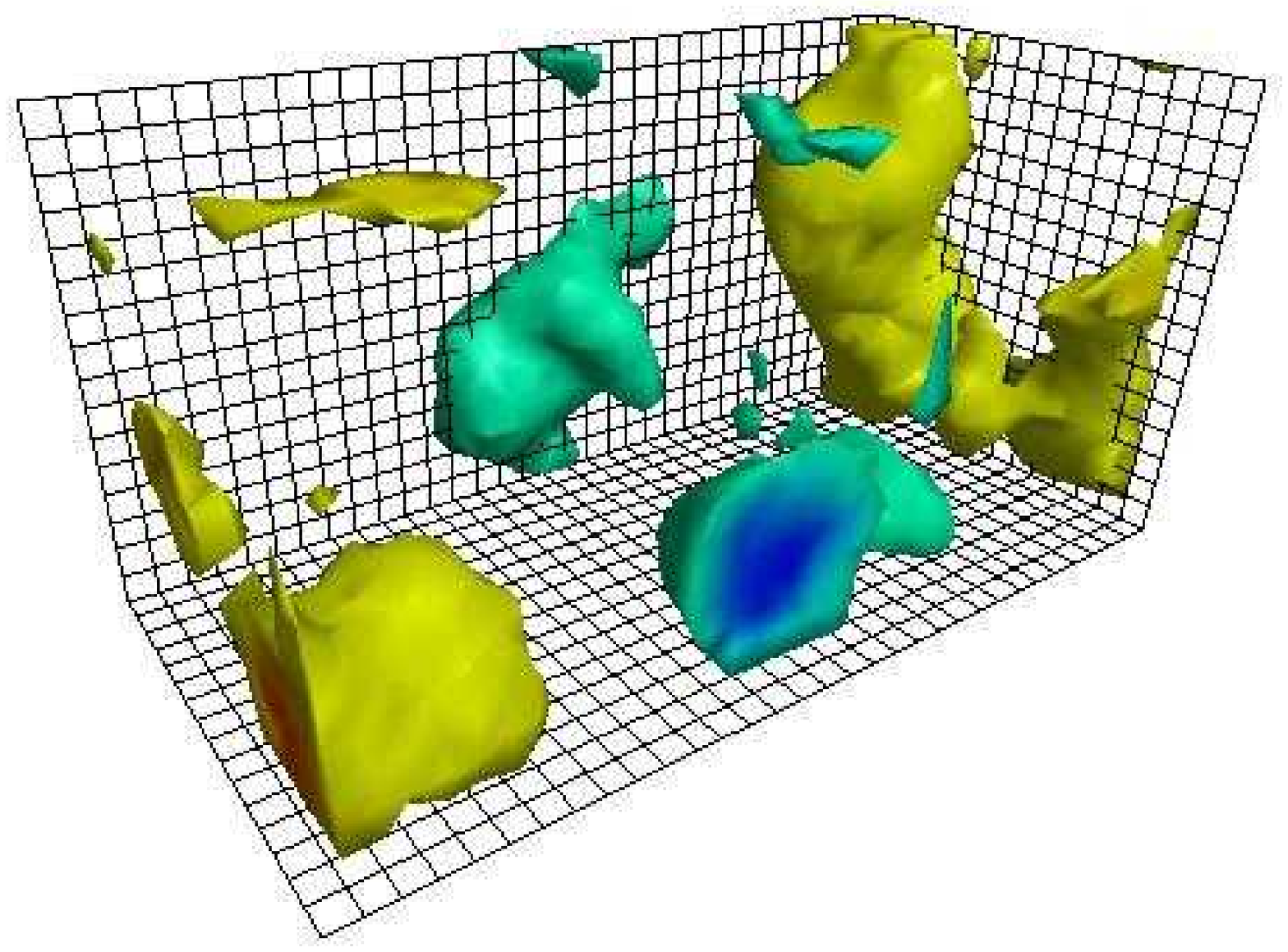} 
      \caption{The scalar (middle) and pseudoscalar (bottom) density
        of the first non-zero mode for the $Q=1$ configuration shown
        previously in Fig.~\ref{fig:zm.0130-rho-0}, along with the
        gluonic topological charge density after 200 sweeps of
        smearing (top). Again, the scalar density extends over objects
        of differing charge, and the regions of alternate charge are
        revealed by the local chirality of the pseudoscalar
        density. In color online: negative density is blue/green,
        positive density is red/yellow (top), positive chirality is
        blue/green and negative chirality is red/yellow (bottom). In
        grey-scale: negative density dark, positive density light
        (top), and positive chirality dark, negative chirality light
        (bottom).}
    \label{fig:zm.0130-rho-1}
    \end{center}
\end{figure*}

\begin{figure*}[!h]
  \begin{center}
      \includegraphics[width=0.54\textwidth,angle=0]{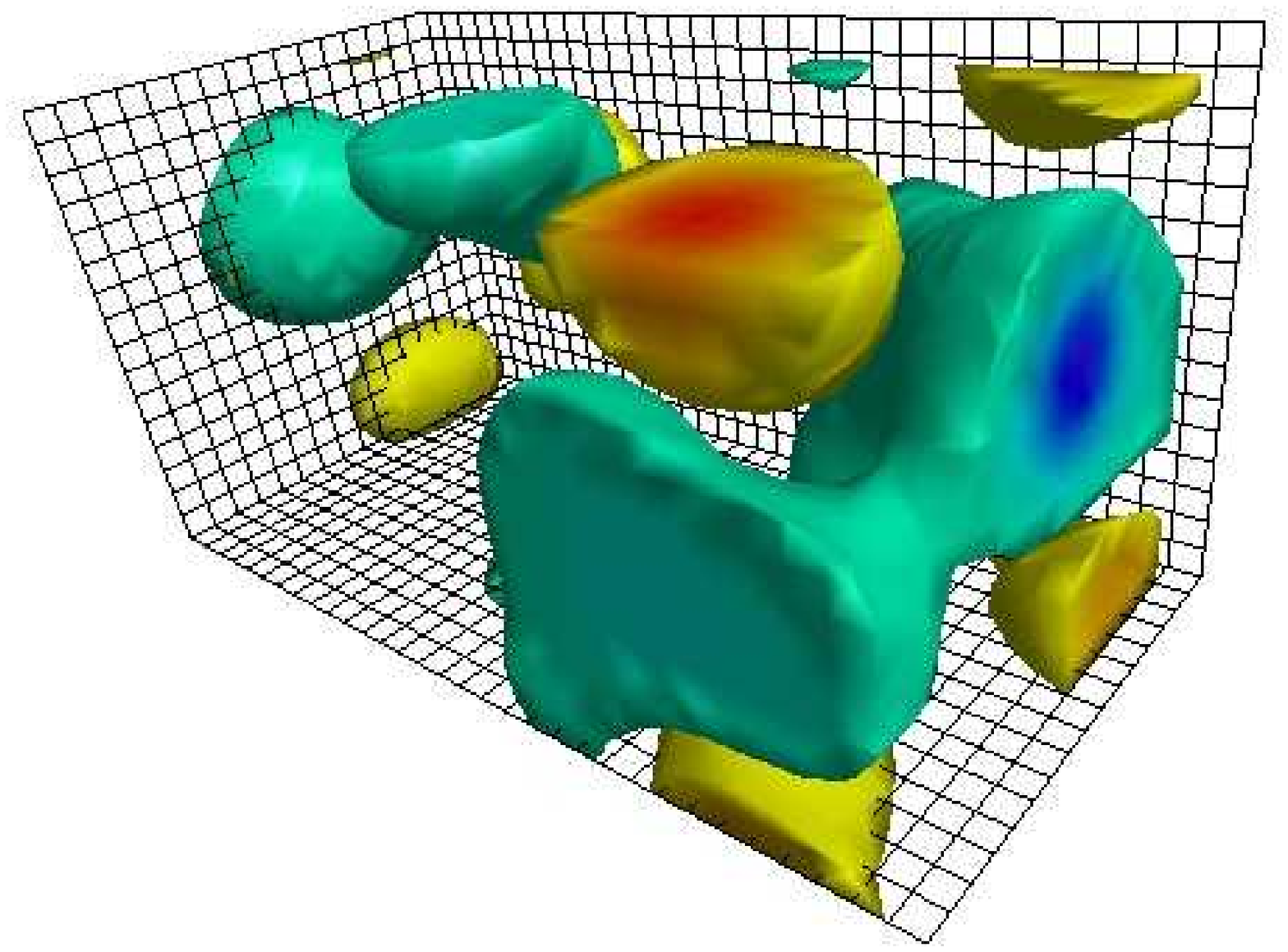} \\
      \hspace{0.5cm}
      \includegraphics[width=0.54\textwidth,angle=0]{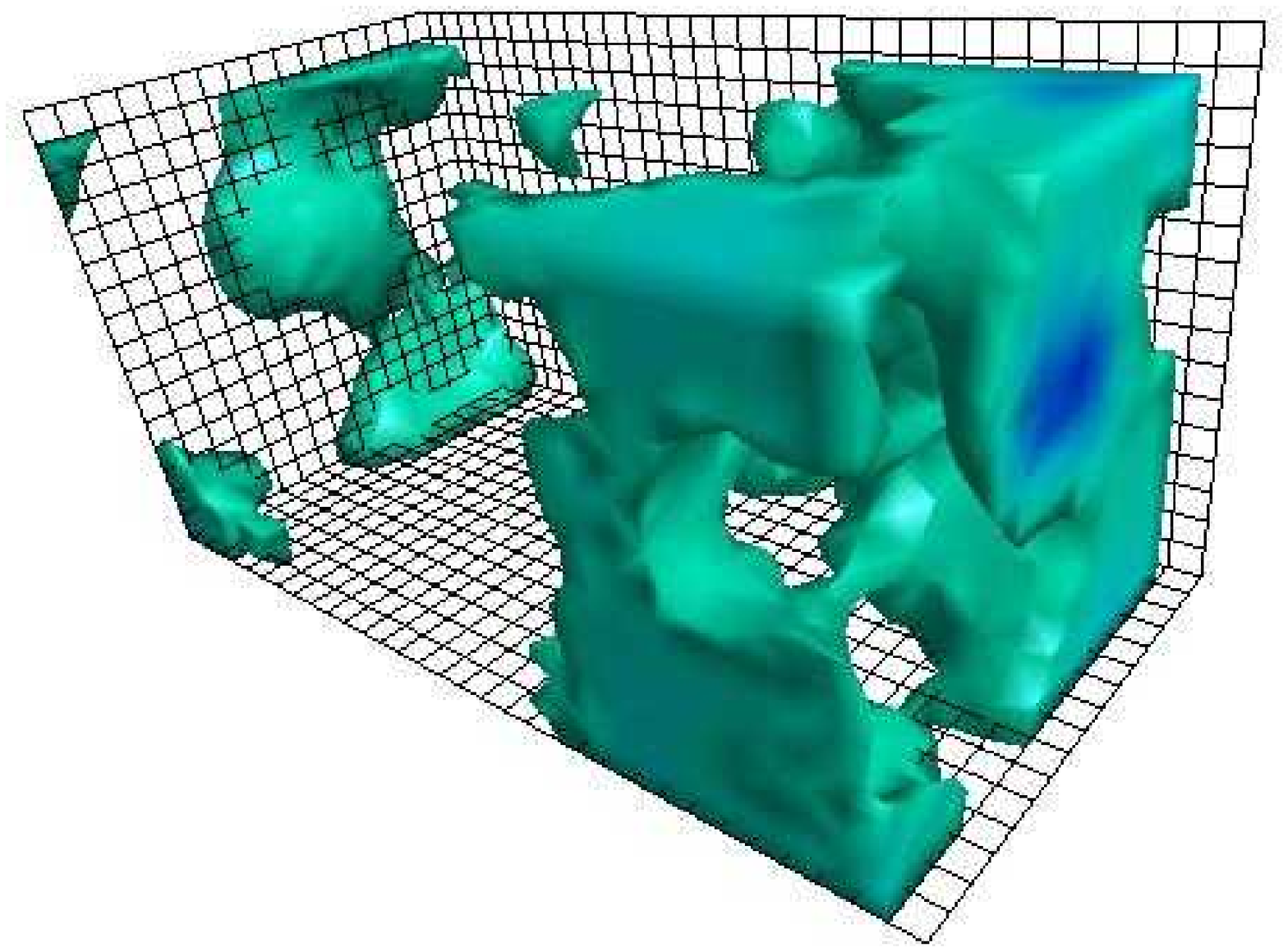} \\
      \hspace{0.5cm}
      \includegraphics[width=0.54\textwidth,angle=0]{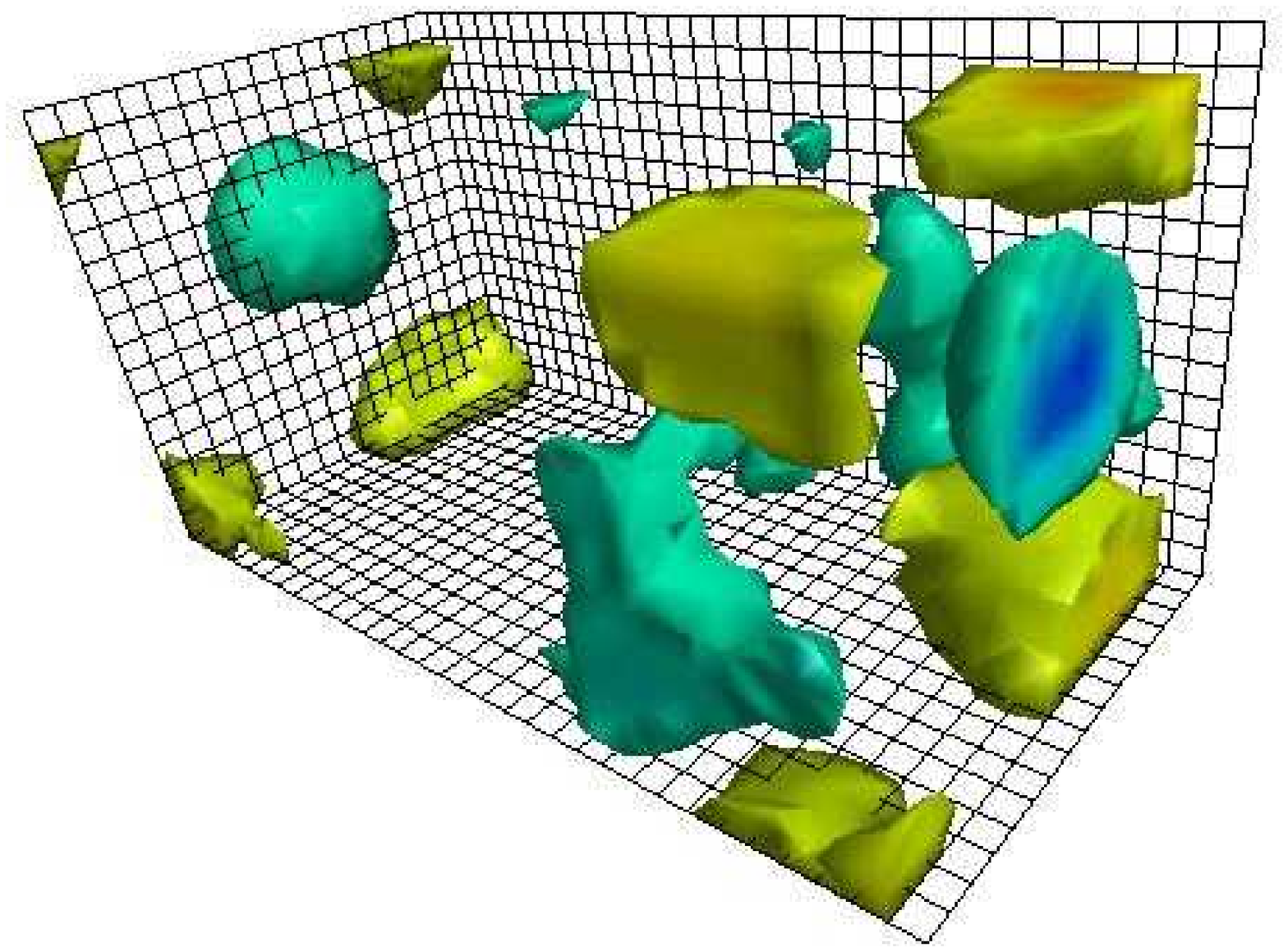}
      \caption{The scalar (middle) and pseudoscalar (bottom) density
        of the first non-zero mode for the $Q=-1$ configuration shown
        previously in Fig.~\ref{fig:zm.0141-rho-0}, along with the
        gluonic topological charge density after 200 sweeps of
        smearing (top). Again, the scalar density extends over objects
        of differing charge, and the regions of alternate charge are
        revealed by the local chirality of the pseudoscalar
        density. In color online: negative density is blue/green,
        positive density is red/yellow (top), positive chirality is
        blue/green and negative chirality is red/yellow (bottom). In
        grey-scale: negative density dark, positive density light
        (top), and positive chirality dark, negative chirality light
        (bottom).}
    \label{fig:zm.0141-rho-1}
  \end{center}
\end{figure*}

The densities for the next two $Q=1$ and $Q=-1$ configurations are
shown in Figs.~\ref{fig:zm.0130-rho-1} and~\ref{fig:zm.0141-rho-1}.
In the instanton model, a whole band of almost-zero modes is generated by 
diagonalizing the Dirac operator in the field of a superposition of 
$n_{+}$ instantons and $n_{-}$ anti-instantons 
in the basis of linear combinations 
of the $n_{+} + n_{-}$ zero modes corresponding to the case of infinite
diluteness. Apart from the remaining $|Q|$ zero modes, the almost-zero modes
are expected to bridge at least one pair of instanton and anti-instanton 
with the scalar density peaking on top of the topological charge lumps and
the pseudoscalar density peaking with the appropriate sign there. 
Qualitatively,
this is visible in Figs.~\ref{fig:zm.0123-rho-1}, 
\ref{fig:zm.0130-rho-1} and \ref{fig:zm.0141-rho-1}.

\section{Conclusions and outlook}
\label{sec:conclusions}
In this paper we have confronted the overlap-fermionic topological charge
density and the improved gluonic topological charge at different levels of
ultraviolet smoothing, realized in one case by a truncation of the mode expansion 
at $\lambda_{cut}$ and in the other case by a certain number of smearing steps
applied to the gauge field in order to wipe out ultraviolet fluctuations. 
These two views of getting the infrared topological structure of the gauge field 
correspond to each other. 
A similar result, however for APE smearing and a different improved Dirac 
operator, has been found
in Ref.~\cite{Bruckmann:2006wf,Solbrig:2007nr}. In the present paper this has 
been confirmed for two other specific realizations of both methods, using the 
massless overlap Dirac operator~\cite{Neuberger:1998wv} on one hand 
and stout-link smearing with respect to an over-improved Symanzik type 
action~\cite{Moran:2007nc}
on the other. These two methods have their respective 
advantages compared to the 
approximate solution~\cite{Gattringer:2000js,Gattringer:2000qu} of the 
Ginsparg-Wilson relation and APE smearing~\cite{Hasenfratz:1999ng}. 

In this paper, the comparison has been 
made more complete and detailed, based first on the density-density two-point
function and second on a point-by-point matching of the respective topological 
densities.
The correspondence between the ultraviolet cutoff $\lambda_{cut}$ of
the overlap analysis and the number of smearing steps justifies the
use of over-improved stout-link smearing, which is computationally
less demanding, in the analysis of topological vacuum structure.
This is of particular interest when investigating the differences
between the vacuum structure of quenched and non-quenched
QCD~\cite{Moran:2007nc,Hasenfratz:1999ng}.

What we did not yet do in the present paper is the natural next step
of formulating
a description of the outstanding clusters of topological charge 
(at each level of ultraviolet smoothing) in terms of an filtered field 
strength, a possibility also offered by the overlap 
analysis~\cite{Ilgenfritz:2007xu}. Furthermore, one should
study the degree of (anti)selfduality of
the gauge field and the percolation properties of the corresponding
(anti)selfdual domains~\cite{Ilgenfritz:2007xu}.
Both questions are accessible not only by the fermionic method but also by
the gluonic method based on stout-link smearing.

\section*{ACKNOWLEDGEMENTS}
E.-M.~I. acknowledges the support from DFG under the grant FOR 465 / Mu932/2-2. 
E.-M.~I., K.~K., G.~S. and V.~W. thank T.~Streuer and Y.~Koma for collaboration
in earlier stages of the overlap project within this DFG-Forschergruppe. 
E.-M.~I. thanks F.~Bruckmann, Ch.~Gattringer, M. M\"uller-Preussker, A.~Sch\"afer 
and S.~Solbrig for collaboration on a similar comparative study within the
DFG-Forschergruppe. 
He is grateful to Lorenz von Smekal for the invitation to Adelaide which was made 
possible by a grant under the Strategic Research Scheme of the Faculty of Sciences 
of the University of Adelaide devoted to the collaboration on Infrared QCD between 
Humboldt University Berlin and Adelaide University.

P.~M. and D.~L. thank the Australian Partnership for Advanced Computing (APAC) 
and the South Australian Partnership for Advanced Computing (SAPAC) for generous 
grants of supercomputer time which have enabled this project. This work is 
supported by the Australian Research Council.


\end{document}